\documentclass[12pt]{article}
\usepackage{a4wide,mathematica,epsf,rotate,float,latexsym,cite}

\newcommand{\cP}{{\cal P}}
\newcommand{\fz}{\footnotesize}
\newcommand{\IM}{\mbox{\rm Im}}
\newcommand{\RE}{\mbox{\rm Re}}
\newcommand{\eqn}[1]{(\ref{#1})}
\newcommand{\mev}{\mbox{\rm MeV}}
\newcommand{\gev}{\mbox{\rm GeV}}
\newcommand{\DeKP}{\Delta_{K\pi}}
\newcommand{\tvs}{\vbox{\vskip 6mm}}
\newcommand{\smvs}{\vbox{\vskip 8mm}}
\newcommand{\bmvs}{\vbox{\vskip 10mm}}
\newcommand{\fgev}{\footnotesize{GeV}}

\newcommand{\newsection}[1]{\section{#1}\setcounter{equation}{0}}

\newcommand{\be}{\begin{equation}}
\newcommand{\ee}{\end{equation}}
\newcommand{\ba}{\begin{eqnarray}}
\newcommand{\ea}{\end{eqnarray}}

\newcommand{\nn}{\nonumber}

\begin{document}

\begin{titlepage}
\begin{flushright}
{\small\sf IFIC/01-26} \\[-1mm]
{\small\sf FTUV/01-1015} \\[-1mm]
{\small\sf HD-THEP-01-10} \\[-1mm]
{\small\sf FZ-IKP(TH)-01-16} \\[12mm]
\end{flushright}

\begin{center}
{\LARGE\bf Strangeness-changing scalar form factors}\\[12mm]

{\normalsize\bf Matthias Jamin${}^{1,*}$, Jos\'e Antonio Oller${}^{2}$
 and Antonio Pich${}^{3}$} \\[4mm]

{\small\sl ${}^{1}$ Institut f\"ur Theoretische Physik, Universit\"at
           Heidelberg,} \\
{\small\sl Philosophenweg 16, D-69120 Heidelberg, Germany}\\
{\small\sl ${}^{2}$ Forschungszentrum J\"ulich, Institut f\"ur
           Kernphysik (Theorie),} \\
{\small\sl D-52425 J\"ulich, Germany}\\
{\small\sl ${}^{3}$ Departament de F\'{\i}sica Te\`orica, IFIC,
           Universitat de Val\`encia -- CSIC,}\\
{\small\sl Apt. Correus 22085, E-46071 Val\`encia, Spain} \\[20mm]
\end{center}

{\bf Abstract:}
We derive expressions for strangeness--changing scalar form factors, which
incorporate known theoretical constraints both at low and high energies. Their
leading behaviour in the limit of a large number of colours is calculated from
the resonance chiral Lagrangian. A set of short-distance constraints on the
scalar resonance couplings is obtained, imposing the form factors to vanish
at infinite momentum transfer. Making use of previous results on S--wave
$K\pi$ scattering \cite{JOP:00a}, and a coupled--channel dispersion--relation
analysis, the $K\pi$, $K\eta$ and $K\eta'$ scalar form factors are calculated
up to 2~GeV. These scalar form factors are a key ingredient in the
determination of the corresponding scalar spectral function which is important
in the extraction of the mass of the strange quark from QCD sum rules as well
as hadronic $\tau$ decays.

\vfill


\phantom{Keywords:} Coupled channels, Final state interactions.

\vspace{4mm}
{\small ${}^{*}$ Heisenberg fellow.}
\end{titlepage}

\newsection{Introduction}

QCD currents are a basic ingredient of the electroweak interactions. A good
understanding of their associated hadronic matrix elements is required to
control the important interplay of QCD in electromagnetic and weak transitions.
In this work we shall investigate a simple, although highly non-trivial
example of such matrix elements, namely strangeness--changing scalar form
factors.

Chiral Perturbation Theory ($\chi$PT) \cite{WE:79,GL:84,GL:85,EC:95,PI:95,ME:93}
provides a very powerful framework to study the low-energy dynamics of the
lightest pseudoscalar octet. The chiral symmetry constraints are stringent
enough to determine the hadronic matrix elements of the light quark currents
at very low energies. Unfortunately, these chiral low-energy theorems only
apply near threshold. To describe the resonance region around 1~GeV, additional
dynamical information is required.

One can construct a chiral--symmetric Effective Field Theory with resonance
fields as explicit degrees of freedom \cite{EGPR:89,EGLPR:89}. Although not as
predictive as the standard chiral Lagrangian for the pseudo-Goldstone mesons,
the resonance chiral Lagrangian \cite{EGPR:89} turns out to provide interesting
results, once additional short-distance dynamical QCD constraints are taken
into account \cite{EGLPR:89}. At tree level, this Lagrangian encodes the
large--$N_c$ properties of QCD \cite{TH:74,WI:79}, in the approximation
\cite{PPR:98} of keeping only the dominant lowest--mass resonance multiplets.

In the following, we present a detailed analysis of the scalar
strangeness--changing form factors, within the resonance chiral framework.
These form factors are needed to improve the strange quark mass determination 
from QCD sum rules for the corresponding scalar current \cite{NPRT:83,JM:95,
ChDPS:95,cps:96,CFNP:97,JA:98,BGM:98,mal:99}. The $\Delta S=1$ scalar form
factors also govern the leading $J=0$ contributions to the Cabibbo--suppressed
hadronic $\tau$ decay width. A good theoretical understanding of these
quantities could result in a more precise determination of $m_s$ from $\tau$
decays \cite{PP:98,ALEPH:99,PP:99,kkp:00,KM:00,ChDGHPP:01}.

In order to work with a renormalisation--group invariant object, we define
the scalar form factors $f_X(s)$ from the divergence of the corresponding
vector current matrix elements,
\begin{equation}
\label{eq:1.1}
\langle 0 |\partial^\mu (\bar s \gamma_\mu u)(0) | X \rangle
\;  = \; i\,(m_s - m_u) \, \langle 0 | (\bar s \, u)(0) | X \rangle
\; \equiv  \; -\,i\,\frac{\Delta_{K\pi}}{\sqrt{2}}\, C_X f_X(s) \,,
\end{equation}
where
$\Delta_{K\pi} \equiv M_K^2 - M_\pi^2$ and $s\equiv p_X^2$. Moreover, we have
factored out the normalisation constants $C_X$, so that $f_X(s) = 1$ at
lowest order in $\chi$PT. Then the $C_X$ for the lowest lying hadronic
systems $K^+\pi^0$, $K^0\pi^+$, $K^+\eta_8$ and $K^+\eta_1$ which contribute
to $f_X(s)$ are given by
\begin{equation}
C_{K^+\pi^0} \; = \; 1 \,, \qquad
C_{K^0\pi^+} \; = \; \sqrt{2} \,, \qquad
C_{K^+\eta_8} \; = \; -{1\over\sqrt{3}} \,, \qquad
C_{K^+\eta_1} \; = \; 2{\sqrt{2\over 3}} \,.
\end{equation}
Below, we shall only work in the isospin limit with $f_{K\pi} \equiv
f_{K^+\pi^0} = f_{K^0\pi^+}$.

The form factors associated with the physical $\eta$ and $\eta'$ fields are
easily obtained, taking into account the mixing between the two isoscalar
states:
\begin{equation}
\label{eq:mixing}
\left(\!\begin{array}{c} \eta \\ \eta' \end{array} \!\right)
\; = \; \left(\! \begin{array}{cc} 
\cos{\theta} & -\sin{\theta} \\ \sin{\theta} & \cos{\theta}
\end{array} \!\right)
\left(\! \begin{array}{c} \eta_8 \\ \eta_1 \end{array} \!\right) \,.
\end{equation}
We shall adopt the normalisation
$C_{K^+\eta} = C_{K^+\eta_8}$ and $C_{K^+\eta'} = C_{K^+\eta_1}$.
Therefore,
\begin{eqnarray}
\label{eq:mixstates}
f_{K\eta}(s) & = & \cos{\theta} \, f_{K\eta_8}(s) + 
2\sqrt{2} \, \sin{\theta} \, f_{K\eta_1}(s) \,, \nn \\
\tvs
f_{K\eta'}(s) & = & \cos{\theta} \, f_{K\eta_1}(s) - 
{1\over 2\sqrt{2}} \, \sin{\theta} \,  f_{K\eta_8}(s) \,.
\end{eqnarray}
For our numerical analyses we shall use $\sin{\theta}=-1/3\approx-20^\circ$,
which is in the ball-park of present phenomenological determinations. Notice
that for this particular value of the mixing angle $f_{K\eta}(s)=0$ to
lowest order in $\chi$PT. Although this cancellation is no longer true at
higher orders in the chiral expansion, it indicates a strong suppression of
the $K\eta$ current matrix element.

We will first derive the theoretical predictions for $f_X(s)$ in the limit of
a large number of colours. This can be achieved in the framework of the
resonance chiral Lagrangian. Imposing these form factors to vanish at infinite
momentum transfer, we obtain a set of short-distance constraints on the chiral
couplings of the scalar meson resonances.

The $\chi$PT loops are subleading corrections in the $1/N_c$ counting. They
incorporate the unitarity field theory constraints, in a perturbative way,
order by order in the momentum expansion. For higher energies around the
resonance region and above, the low-energy expansion breaks down because
the unitarity corrections due to re-scattering are no longer perturbative.
Therefore some kind of resummation of those chiral loops is required to satisfy
unitarity \cite{tru:88,dht:90,oo:97,GP:97,GU:98,OOP:99,OO:99,OOR:99,dpp:00,
mo:00,oop:00,pp:01}. This effect appears to be crucial for a correct
understanding of the scalar sector, because the S--wave re-scattering of two
pseudoscalars is very strong.

In a previous paper \cite{JOP:00a}, we have presented a detailed study of
S--wave $K\pi$ scattering up to $2\;\gev$, within the same resonance chiral
framework supplemented with the unitarisation procedure developed in refs.
\cite{OO:99,oll:00}. We will use those results to perform a calculation of the
scalar form factors from dispersion relations, explicitly obeying unitarity,
also taking into account coupled--channel effects.

In section~2 first the scalar form factors are calculated at tree level in
the $\chi$PT framework including resonances. Then their expressions in
conventional $\chi$PT at the next-to-leading order are reviewed and compared
to the resonance $\chi$PT approach. In section~3 we derive constraints on
the scalar form factors by exploiting unitarity and analyticity. These
constraints result in a set of coupled dispersion integral equations for the
form factors. The Omn\`es solution of the dispersion relation in the elastic
single channel case is also discussed. Section~4 contains our numerical
analysis for the single as well as coupled--channel cases and our central
results for the scalar form factors are presented. Finally we close with
some conclusions and an outlook in section~5.

\newsection{Effective Lagrangian Results}

\subsection{Resonance Chiral Lagrangian}

In the limit of an infinite number of quark colours, QCD reduces to a theory
of tree--level resonance exchanges \cite{TH:74,WI:79}. At low energies, the
dominant effects are governed by the lowest--mass meson multiplets, and can be
analysed within the resonance chiral Lagrangian framework developed in refs.
\cite{EGPR:89,EGLPR:89}. We refer to those references for details on the chiral
formalism and notations. We will use a ${\rm U}(3)_L\!\times\!{\rm U}(3)_R$
effective theory with nine pseudoscalar Goldstone bosons, which is the
appropriate framework in the limit $N_c$ to infinity \cite{leu:96,her:97,
her:98}.

The tree--level calculation of the relevant scalar form factors is
straightforward. One obtains:
\begin{eqnarray}
\label{eq:ResFF}
f_{K\pi}(s) & = & 1 + {4 c_m\over f^2} \biggl[\, c_d + (c_m - c_d)\,
{M_K^2+M_\pi^2 \over M_S^2}\,\biggr] {s \over M_S^2 -s} \,, \nn \\
\smvs
f_{K\eta_8}(s) & = & 1 + {4 c_m\over f^2 (M_S^2-s)}\biggl[\,c_d \Big(s-M_K^2-
p^2_{\eta_8}\Big) + c_m \Big(5 M_K^2-3 M_\pi^2 \Big)\,\biggr] \nn \\
\smvs && \hspace{3mm}
+\,{4 c_m(c_m - c_d)\over f^2 M_S^2}\,\Big( 3 M_K^2- 5 M_\pi^2 \Big) \,, \\
\smvs
f_{K\eta_1}(s) & = & 1 + {4 c_m\over f^2 (M_S^2-s)}\biggl[\,c_d \Big(s-M_K^2-
p^2_{\eta_1}\Big) + c_m\, 2 M_K^2\,\biggr] - {4 c_m(c_m - c_d)\over f^2 M_S^2}
\, 2 M_\pi^2 \,, \nn
\end{eqnarray}
where $c_d$ and $c_m$ are the couplings of the scalar multiplet to the
Goldstone bosons, in the lowest--order chiral resonance Lagrangian of
refs.~\cite{EGPR:89,EGLPR:89}, $M_S$ is the scalar resonance mass and
$f\approx f_\pi = 92.4\;\mev$ is the pion decay constant. The iso-singlet
momentum dependence $p^2_{\eta_i}$ refers to the appropriate meson mass
squared in the linear combinations (\ref{eq:mixstates}), associated with
the physical $\eta$ and $\eta'$ fields.

The numerical values of the couplings $c_d$ and $c_m$ are not very well known.
We can get a theoretical constraint by enforcing the scalar form factors to
vanish at large momentum transfer. This seems a very reasonable
phenomenological assumption for a composite object. All three form factors
are zero when $s$ goes to infinity, if the following two conditions are
satisfied:
\begin{equation}
\label{cdm_constraints1}
4 c_d c_m \; = \; f^2 \,, \qquad\qquad  c_m - c_d \; = \; 0\,.
\end{equation}
This implies
\begin{equation}
\label{cdcm1}
c_d \; = \; c_m \; = \; f/2 \; \approx \; 46\;\mev \,,
\end{equation}
and a dipole structure for the scalar form factors,
\begin{equation}
\label{eq:dipole}
f_X(s) \; = \; { f_X(0)\over (1 - s/M_S^2)} \,.
\end{equation}
Taking $\sin{\theta} = -1/3$ and $M_S = M_{K^*_0}\approx 1.4$ GeV,
\begin{eqnarray}
\label{F3ChPT}
f_{K\pi}(0) &=& 1 \, , \qquad\qquad
f_{K\eta}(0) \; = \; 2\sqrt{2}\;{\Delta_{K\pi}\over M_{K^*_0}^2}
\; \approx \; 0.33 \,, \nn \\
\smvs
f_{K\eta'}(0) &=& {1 \over 2\sqrt{2}} \biggl[\, 3 +
{3(M_K^2-M_{\eta'}^2) + \Delta_{K\pi}\over M_{K^*_0}^2}\,\biggr]
\; \approx \; 0.74 \,.
\end{eqnarray}
Notice that $f_{K\eta}(0)$ goes to zero in the limit $M_K = M_\pi$.

The above results can easily be generalised to take into account the
exchange of $N$ different scalar multiplets with parameters $M_{S,i}$,
$c_{d,i}$ and $c_{m,i}$ \ ($i=1,\cdots,N$). In that case, $f_X(s)-1$ contains
a sum of $N$ contributions like the ones in eqs.~(\ref{eq:ResFF}), with the
appropriate changes on the scalar parameters. The short--distance requirement
that $f_X(s)$ goes to zero for $s\to\infty$ then implies the constraints:
\begin{equation}
\label{cdm_constraints2}
4 \sum_{i=1}^N c_{d,i} c_{m,i} \; = \; f^2 \,, \qquad\qquad  
\sum_{i=1}^N {c_{m,i}\over M_{S_i}^2}  \left(c_{m,i} - c_{d,i}\right) 
\; = \; 0 \,.
\end{equation}
The consequences of these conditions have already been investigated in the
analysis of S--wave $K\pi$ scattering, performed in ref.~\cite{JOP:00a}. A
more detailed discussion of the phenomenological implications of the relations
\eqn{cdm_constraints2} in the case of two scalar resonances will be given at
the end of the next section.

\subsection{Chiral Perturbation Theory}

The $K\pi$ and $K\eta_8$ scalar form factors have been computed at the one-loop
level in ${\rm SU}(3)_L\!\times\!{\rm SU}(3)_R$ $\chi$PT \cite{GL:85}. The
result can be written as:
\begin{eqnarray}
\label{ffxpt}
f_{K\pi}(s)&=& 1 +\frac{4 L_5^r(\mu)}{f^2}s +  \frac{1}{8 f^2}
\left(5s-2\Sigma_{K\pi}-3\frac{\Delta^2_{K\pi}}{s} \right)
\bar{J}_{K\pi}(s) 
\nonumber \\ &&
\!\!\!\!\!\!\!\!\!\!\!\!\!\!\!\!\!\!\!\!\!\!\!\!\!\!\!\!\!\!\mbox{}
+\frac{1}{24 f^2}\left(3s-2\Sigma_{K\pi}-\frac{\Delta^2_{K\pi}}{s} \right)
\bar{J}_{K\eta_8}(s) 
+ \frac{s}{4\Delta_{K\pi}}\left(5 \mu_\pi-2\mu_K-3\mu_{\eta_8} \right)\,, \\
\bmvs
\label{ffeta_xpt}
f_{K\eta_8}(s)&=& 3 {M_{\eta_8}^2 - M_K^2 \over \Delta_{K\pi}} \left\{
1 +\frac{4 L_5^r(\mu)}{f^2}s +  \frac{3}{8 f^2}
\left(3s-2\Sigma_{K\pi}-\frac{\Delta^2_{K\pi}}{s} \right)
\bar{J}_{K\pi}(s) 
\right.\nonumber \\ &&
\!\!\!\!\!\!\!\!\!\!\!\!\!\!\!\!\!\!\!\!\!\!\!\!\!\!\!\!\!\!
\left.\mbox{}
-\frac{1}{24 f^2}\left(9s-2 M^2_K - 18 M_{\eta_8}^2
+\frac{\Delta^2_{K\pi}}{s} \right) \bar{J}_{K\eta_8}(s) 
+ \frac{9s}{4\Delta_{K\pi}}\left( \mu_\pi-2\mu_K+\mu_{\eta_8} \right)
\right\} \,,
\end{eqnarray}
where $\Sigma_{K\pi}\equiv M_K^2 + M_\pi^2$.

The unitarity corrections associated with the re-scattering of the final
pseudoscalar particles are incorporated through the loop functions
$\bar J_{PQ}(s)$. Their explicit expressions are given in appendix~A.
Although the contributions from Goldstone loops are next-to-leading in the
$1/N_c$ expansion, they generate large logarithmic corrections to the scalar
form factors. The loops introduce a dependence on the renormalisation scale
$\mu$ in the renormalised chiral coupling $L_5^r(\mu)$ and through the
explicit factors
\begin{equation}
\mu_P \; \equiv \; {M_P^2 \over 32\pi^2 f^2} \ln{\left(M_P^2/\mu^2\right)} \,.
\end{equation}
Nevertheless, the scalar form factors are of course independent of $\mu$.

In the large--$N_c$ limit, the expressions for the form factors $f_{K\pi}(s)$
and $f_{K\eta_8}(s)$ reduce to
\begin{equation}
\label{eq:ffp4Nc}
f_{K\pi}(s) \; = \; {\Delta_{K\pi}\over 3 (M_{\eta_8}^2 - M_K^2)}\;
f_{K\eta_8}(s) \; = \; 1 + {4 L_5 s\over f^2} \,.
\end{equation}
The comparison\footnote{
The $K\eta_8$ scalar form factor in \eqn{eq:ResFF} includes an additional term
proportional to $c_m c_d \Delta_{K\pi}/M_S^2$, which is generated through
$\eta_8$--$\eta_1$ mixing.}
with the resonance--exchange results in eqs.~\eqn{eq:ResFF} gives the
corresponding scalar contribution to the ${\cal O}(p^4)$ chiral coupling $L_5$:
\begin{equation}
L_5^S \;=\; {c_m c_d \over M_S^2} \;\approx\; {f^2\over 4 M_S^2} \,.
\end{equation}
The resonance propagators appearing in \eqn{eq:ResFF} provide an explicit
resummation of local terms to all orders in the chiral expansion, improving
the result \eqn{eq:ffp4Nc} to the dipole form \eqn{eq:dipole}. Adding a
second scalar resonance with mass $M_{S'}^2$ and couplings $c_d'$ and $c_m'$,
and also taking into consideration the ${\cal O}(p^4)$ coupling constant $L_8$,
the contributions from the scalar resonances take the form \cite{EGPR:89}:
\begin{equation}
\label{L5SL8S}
L_5^S \;=\; \frac{c_d c_m}{M_S^2} + \frac{c_d' c_m'}{M_{S'}^2}
\qquad \mbox{and} \qquad
L_8^S \;=\; \frac{c_m^2}{2M_S^2} + \frac{c_m'^2}{2M_{S'}^2} \,.
\end{equation}

Now, we are in a position to discuss phenomenological consequences of the above
relations for the scalar couplings. Using the second of the short--distance
constraints \eqn{cdm_constraints2}, together with the relations \eqn{L5SL8S},
immediately implies $L_5^S=2L_8^S$, independent of the number of scalar
resonances. Within the uncertainties, this relation is fulfilled by the most
recent determination of the chiral couplings $L_5$ and $L_8$ \cite{abt:01},
\begin{equation}
\label{L5L8}
L_5 \;=\; (0.91 \pm 0.15)\cdot 10^{-3}
\qquad \mbox{and} \qquad
L_8 \;=\; (0.62 \pm 0.20)\cdot 10^{-3} \,,
\end{equation}
suggesting that both the resonance saturation as well as the short--distance
constraints are reasonable approximations. However, we only have three linearly
independent relations for the four scalar couplings $c_d$, $c_m$, $c_d'$ and
$c_m'$, so that we have to make further assumptions to obtain estimates for
these couplings.

The estimate \eqn{cdcm1} with only one scalar resonance led to $c_d=c_m$ and
thus it seems plausible to keep this constraint in addition. One immediate
consequence of the constraint and the second relation \eqn{cdm_constraints2}
is $c_d'=c_m'$. Employing the first of the short--distance constraints
\eqn{cdm_constraints2} and the relations \eqn{L5SL8S}, the remaining couplings
$c_d$ and $c_d'$ can be calculated with the result:
\begin{equation}
\label{cdcdp}
c_d  \;=\; c_m  \;=\; 37.0 \; \mev
\qquad \mbox{and} \qquad
c_d' \;=\; c_m' \;=\; 27.7 \; \mev \,.
\end{equation}
Here, the phenomenological value \eqn{L5L8} for $L_5$, as well as $M_S=1.4\;
\gev$ and $M_S'=1.9\;\gev$ have been used. The approach of utilising the
short--distance constraints, together with $c_d=c_m$ has also been followed
in our fit (6.10) of ref.~\cite{JOP:00a}. The fit then resulted in 
\begin{equation}
\label{cdcdp6.10}
c_d  \;=\; c_m  \;=\; 23.8 \; \mev
\qquad \mbox{and} \qquad
c_d' \;=\; c_m' \;=\; 39.6 \; \mev \,.
\end{equation}
These values would correspond to $L_5^S=0.72\cdot 10^{-3}$, which in view of
the phenomenological result \eqn{L5L8} appears a little low, although
nevertheless acceptable. One should note that at the tree--level there is
some ambiguity in the scalar resonance mass $M_S$, and e.g. taking $M_S=1.2\;
\gev$, eq.~\eqn{cdcdp} would change to $c_d=27.6\;\mev$ and $c_d'=37.0\;\mev$,
much closer to the result \eqn{cdcdp6.10}.

A different, although also not implausible assumption would be universality of
the chiral couplings, namely $c_d=c_d'$ and $c_m=c_m'$. These constraints allow
us to calculate $L_5^S$ from the first of the short--distance relations:
\begin{equation}
\label{L5S}
L_5^S \;=\; \frac{f^2}{8}\,\biggl(\frac{1}{M_S^2}+\frac{1}{M_{S'}^2}\biggr)
\;=\; 0.84\cdot 10^{-3} \,,
\end{equation}
surprisingly consistent with the phenomenological value \eqn{L5L8}. On the
other hand, just using resonance saturation of the chiral couplings
\eqn{L5SL8S} and their values \eqn{L5L8}, yields
\begin{equation}
\label{cdcm}
c_d \;=\; c_d' \;=\; 29.1 \; \mev
\qquad \mbox{and} \qquad
c_m \;=\; c_m' \;=\; 39.7 \; \mev \,,
\end{equation}
which also satisfies the first short--distance relation rather well. However,
these parameters violate the second of the short--distance constraints.
Enforcing the second constraint in addition again requires $c_d=c_m$, and we
fall back to the first assumption. The values of the chiral scalar couplings
which correspond to this last scenario are
\begin{equation}
\label{cdcmcdpcmp}
c_d  \;=\; c_m  \;=\; c_d' \;=\; c_m' \;=\; 32.7 \; \mev \,.
\end{equation}
The ranges of the scalar couplings which arose in the discussion above should
give some indication of their present uncertainties which we conclude to be
about 30\%.

\newsection{Analyticity and unitarity constraints}

In this section, we present the formalism employed to calculate the coupled
channel $K\pi$, $K\eta$ and $K\eta'$ scalar form factors up to around 2 GeV.
The $\chi$PT results \eqn{ffxpt} and \eqn{ffeta_xpt} are only valid at low
momentum transfers. We have been able to resum the local leading (in $1/N_c$)
contributions to all chiral orders, through the resonance propagators.
Analyticity and unitarity allow us to obtain further constraints on the scalar
form factors. We shall first derive unitarity relations obeyed by the scalar
form factors which link these quantities to the S--wave $I=1/2$ $K\pi$,
$K\eta$ and $K\eta'$ partial wave amplitudes. These amplitudes will be taken
from our previous work~\cite{JOP:00a}, where they were studied in detail up to
$2\,\gev$ in the framework of $\chi$PT with resonances, supplemented with a
suitable unitarisation procedure.

\subsection{Unitarity relations}

Unitarity of the scattering matrix $S$ immediately implies the following
identity for the T--operator,
\begin{equation}
\label{uni3}
{T}-{T}^\dagger \; = \;i\,{T}\cdot {T}^\dagger \,,
\end{equation}
where $T$ is defined by $S\equiv 1+i\,T$. To obtain the form factors in
question, the previous general relation is applied to the transition of the
states $|K\phi_k\rangle$ (where $\phi_1=\pi$, $\phi_2=\eta$ and $\phi_3=\eta'$)
to the vacuum state. The scalar operator giving rise to this transition should
have strangeness $|S|=1$ and isospin $I=1/2$. Let us remark that we are
considering a pure strong interaction problem without electroweak corrections.

Inserting a complete set of intermediate states on the right--hand side of
eq.~\eqn{uni3}, and restricting this set to the two--particle states
$|K\phi_i\rangle$, we arrive at
\begin{equation}
\label{delta3}
\langle 0|T|K\phi_k\rangle-\langle 0|T^\dagger|K\phi_k\rangle \; = \;
i\sum_i\,\frac{\theta(s-s_{th\,i})\,q_i(s)}{8\pi \sqrt{s}} \;\langle 0|T|
K\phi_i\rangle \!\int \langle K\phi_i|T^\dagger|K\phi_k\rangle\,d\cos\theta \,,
\end{equation}
where $s=p_i^2$ with $p_i$ being the total four--momentum of the state
$|K\phi_i\rangle$, $q_i\equiv\lambda_{K\phi_i}/(2\sqrt{s})$ is the modulus
of the centre-of-mass three--momentum for this state, and $\theta$ is the
corresponding scattering angle. The function $\lambda_{K\phi_i}$ is defined
in appendix~A, and the trivial centre-of-mass motion has been removed.

Performing a partial wave decomposition of $T^{ik}(s,cos\,\theta)\equiv
\langle K\phi_i|T|K\phi_k\rangle$, we have:
\begin{equation}
\label{ppw}
T^{ik}(s,z) \; = \; 16\pi \sum_{l=0}^\infty \,(2l+1)\,
t_l^{ik}(s)\,{\mathcal{P}}_l(z) \,,
\end{equation}
where ${\mathcal{P}}_l(z)$ is the Legendre polynomial of degree $l$ with $l$
being the orbital angular momentum and $t_l^{ik}(s)$ are the partial
wave amplitudes. Making use of time reversal invariance and inserting the
decomposition \eqn{ppw} into eq.~\eqn{delta3}, only the S--wave component
survives the angular integration. We thus obtain our central unitarity relation
for the scalar form factors $F_k(s)\equiv \langle 0|T|K\phi_k\rangle$:
\begin{equation}
\label{unif}
\IM \,F_k(s) \; = \; \sum_i \,\sigma_i(s) F_i(s) \,
t_0^{ik}(s)^* \,,
\end{equation}
where, for convenience, we have defined the quantity $\sigma_i(s) \equiv
\theta(s-s_{th\,i})\,2q_i(s)/\sqrt{s}$. For an appropriate choice of the
scalar T--operator with isospin $1/2$, in accordance with eq.~\eqn{eq:1.1},
the form factors $F_k(s)$ are related to $f_X(s)$ of section~1 by:
\begin{equation}
F_{K\pi}(s) \; = \; f_{K\pi}(s) \,, \qquad
F_{K\eta}(s) \; = \; \frac{C_{K\eta}}{\sqrt{3}}\,f_{K\eta}(s) \,, \qquad
F_{K\eta'}(s) \; = \; \frac{C_{K\eta'}}{\sqrt{3}}\,f_{K\eta'}(s) \,.
\end{equation}
We use an isospin basis of states $|K\phi_k\rangle$. The coupling of the
$|K\pi\rangle$ state with $I=I_3=1/2$ to the scalar current in (1.1) is
a factor $\sqrt{3}$ larger than the one of $|K^+\pi^0\rangle$. Too keep the
same normalisation as for $f_{K^+\pi^0}(s)$, we rescale all form factors by
a global factor $1/\sqrt{3}$.

As stated above, the sum in eq.~\eqn{unif} is restricted to two--particle
intermediate states. Thus some comments about the role of multiparticle states
are in order. The lightest multiparticle state contributing to this sum is
$|K\pi\pi\pi\rangle$. From the theoretical side, its contributions are 
suppressed both in the chiral and large--$N_c$ expansions. Experimentally, in
\cite{est:78,ast:88} it has been established that the $I=1/2$ S--wave $K\pi$
amplitude is elastic below roughly $1.3\,\gev$ with $K\eta'$ being the first
relevant inelastic channel. These conclusions have been confirmed in
\cite{JOP:00a} where it was demonstrated that the contribution of the $K\eta$
channel to the S--wave amplitude could be neglected. However, it was also found
in this work that for energies higher than about $2\,\gev$ other inelastic
channels become increasingly important.

The unitarity relation \eqn{unif} given above poses tight constraints on the
scalar form factors. For the elastic case one simply has:
\begin{equation}
\label{uniba1}
\IM\, F_1(s) \; = \; \sigma_1(s) F_1(s) \, t_0^{11}(s)^*
\end{equation}
which implies the well known Watson final state theorem \cite{WA:55}, stating
that the phase of $F_1(s)$ is the same as the one of $t_0^{11}(s)$. This is
obvious from the previous equation since its left--hand side is real.

Let us now discuss the two--channel case, which is very appropriate since the
contributions from the $K\eta$ state are negligible to a very good
approximation. Substituting $\IM\,F_k(s)$ by $(F_k(s)-F_k(s)^*)/2i$ in
eq.(\ref{unif}), one finds above the threshold of the $K\eta'$:
\begin{eqnarray}
\label{uniba21}
F_1(s) &=& \Big(1+2i\,\sigma_1(s)\,t_0^{11}(s)\Big) F_1(s)^*+
           2i\,\sigma_3(s)\,t_0^{13}(s)\,F_3(s)^* \,, \\
\tvs
\label{uniba22}
F_3(s) &=& 2i\,\sigma_1(s)\,t_0^{13}(s)\,F_1(s)^*+
           \Big(1+2i\,\sigma_3(s)\,t_0^{33}(s)\Big) F_3(s)^* \,,
\end{eqnarray}
where, although working with two channels, we have used the label 3 to
indicate the $K\eta'$ state as introduced before. Following ref.\cite{babelon},
we now express $F_3(s)$ in terms of $F_1(s)$ from eqs.~\eqn{uniba21} and
\eqn{uniba22}.

The partial wave amplitudes $t_0^{mn}$ can be parametrised by introducing the
symmetric and unitary $2\times 2$ S--matrix $S_{mn}=\delta_{mn} +
2i\sqrt{\sigma_m \sigma_n}\,t_0^{mn}$:
\begin{equation}
\label{S}
S \; = \; \left(\! 
\begin{array}{cc}
\eta\,\exp 2i\delta_1(s) & i\sqrt{1-\eta^2}\exp i(\delta_1(s)+\delta_3(s)) \\
i\sqrt{1-\eta^2}\exp i(\delta_1(s)+\delta_3(s)) & \eta\,\exp 2i\delta_3(s)
\end{array}
\!\right)
\end{equation}
with $\eta\equiv \cos 2\alpha$ ($\sin2\alpha=\sqrt{1-\eta^2}$) the inelasticity
parameter ($0\leq\eta\leq 1)$ and $\delta_k(s)$ the phase shift of channel $k$.
We further express the form factors as $F_k(s)=f_k \exp i(\delta_k+\phi_k)$
with $f_k$ and $\phi_k$ real functions and $f_k\geq 0$. Taking real and
imaginary parts in eqs.~\eqn{uniba21} and \eqn{uniba22}, the following set
of relations emerges:
\begin{eqnarray}
\label{uniba31}
(1-\cos 2\alpha) \cos \phi_1 f_1 &=& \sqrt{\frac{\sigma_3}{\sigma_1}}\,
\sin 2\alpha \, \sin \phi_3 f_3 \,, \\
\tvs
\label{uniba32}
(1+\cos 2\alpha) \sin \phi_1 f_1 &=& \sqrt{\frac{\sigma_3}{\sigma_1}}\,
\sin 2\alpha \, \cos \phi_3 f_3 \,;
\end{eqnarray}
plus two analogous relations with the labels 1 and 3 exchanged. The latter
relations are, however, linearly dependent to the first two. Dividing the
two eqs.~\eqn{uniba31} and \eqn{uniba32} results in:
\begin{equation}
\label{dfase}
\tan \phi_1 \tan\phi_3 \; = \; \tan^2 \!\alpha \,,
\end{equation}
whereas adding them in quadrature yields:
\begin{equation}
\label{mod}
\frac{\sigma_3 f_3^2}{\sigma_1 f_1^2} \; = \; \tan^2\!\alpha +
(\cot^2\!\alpha-\tan^2\!\alpha)\sin^2\phi_1 \,.
\end{equation}
Thus, once $f_1$ and $\phi_1$ are known, $f_3$ and $\phi_3$ can be calculated
from eqs.~\eqn{dfase} and \eqn{mod}. The treatment in the full three--channel
case will be discussed further below.

\subsection{Dispersion relations}

The scalar form factors $F_k(s)$ are analytic functions in the complex
$s$--plane, except for a cut along the positive real axis, starting at the
first physical threshold $s_{th\,1} = (M_K+M_\pi)^2$, where their imaginary
parts develop discontinuities. They are real for $s < s_{th\,1}$. As should
be clear from eq.~\eqn{unif}, their imaginary parts just correspond to the
contributions from all possible on-shell intermediate states.\footnote{We are
excluding the presence of bound state poles below the threshold of the $K\pi$
state.}

Analyticity relates the real and imaginary parts of $F_k(s)$ through
dispersion relations:
\begin{equation}
\label{dis}
F_k(s) \; = \; \frac{1}{\pi} \int\limits^\infty_{s_{th\,1}}\!\! ds'\,
\frac{\IM F_k(s')}{(s'-s-i0)} + {\rm subtractions} \,.
\end{equation}
From unitarity, as shown above, $\IM F_k(s)$ takes the form of eq.~\eqn{unif}.
With the reasonable assumption that for $s$ going to infinity, $F_k(s)$
vanishes sufficiently fast, the form factors satisfy dispersion relations
without subtractions. This assumption was already explored in section~2.

Because $K\pi$ scattering is basically elastic up to around $1.3\,\gev$, it
is again instructive to consider the single channel case. Then the dispersion
relation takes the form:
\begin{equation}
\label{dis1}
F_1(s) \; = \; \frac{1}{\pi}\int\limits^\infty_{s_{th\,1}}\!\!ds'\,
\frac{\sigma_1(s')F_1(s')\,t_0^{11}(s')^*}{(s'-s-i0)} \,.
\end{equation}
In this case the partial wave amplitude can be expressed as
$\sigma_1 t_0^{11}=\,\sin\delta_1\exp(i\delta_1)$, and the previous
eq.~\eqn{dis1} admits the well known Omn\`es--exponential solution
\cite{om:58}:
\begin{equation}
\label{omel}
F_1(s) \, = \, P(s) \exp \left( \frac{s}{\pi} \int\limits^\infty_{s_{th\,1}}
\!\!ds'\, \frac{\delta_1(s')}{s'(s'-s-i0)}\right) \,, 
\end{equation}
with $P(s)$ being a real polynomial to take care of the zeros of $F_1(s)$ for
finite $s$ and $\delta_1(s)$ is the S--wave $I=1/2$ elastic $K\pi$ phase shift.

In writing eq.~\eqn{omel}, we have included one subtraction at the origin since
generally $\delta_1(s)$ tends to a constant for $s$ going to infinity.
A general solution, valid for any number $n$ of subtractions at an arbitrary
subtraction point $s_0$, has been given in refs.~\cite{GP:97,PP:00}, where
the equivalence of the different expressions has been discussed in detail.
The form factor $F_1(s)$ in the elastic case will be discussed further in
our numerical analysis below.

In order to implement the dispersion relation for coupled channels, we neglect,
for the moment, the $K\eta$ contribution and only consider the more important
$K\pi$ and $K\eta'$ channels. For this case, two coupled integral equations
arise:
\begin{eqnarray}
\label{2in}
F_1(s) &=& \frac{1}{\pi}\int\limits_{s_{th\,1}}^\infty \!\!ds'\,
\frac{\sigma_1(s') F_1(s')\,t_0^{11}(s')^*}{(s'-s-i0)} +
\frac{1}{\pi}\int\limits_{s_{th\,3}}^\infty \!\!ds'\,
\frac{\sigma_3(s') F_3(s')\,t_0^{13}(s')^*}{(s'-s-i0)} \,, \nn \\
\smvs
F_3(s) &=& \frac{1}{\pi}\int\limits_{s_{th\,1}}^\infty \!\!ds'\,
\frac{\sigma_1(s') F_1(s')\,t_0^{13}(s')^*}{(s'-s-i0)} +
\frac{1}{\pi}\int\limits_{s_{th\,3}}^\infty \!\!ds'\,
\frac{\sigma_3(s') F_3(s')\,t_0^{33}(s')^*}{(s'-s-i0)} \,.
\end{eqnarray}
The integral equations \eqn{2in} will be solved iteratively according to a
procedure already applied in ref.~\cite{dona:90}. In the first step initial
functions $F_k^{(0)}(s)$ for the form factors, to be specified further below,
are inserted in the right--hand side of eqs.~\eqn{2in}, and new resulting form
factors $F_k^{(1)}(s)$ are calculated. Then the procedure is iterated until
after $n$ steps the procedure converges and the final scalar form factors
$F_k^{(n)}(s)$ are obtained.

The dispersion relations for the three--channel case including the $K\eta$
contribution are completely analogous to the eqs.~\eqn{2in}. Thus the
explicit expressions have been relegated to appendix~B. As for the two--channel
case, they are solved iteratively. Further details on the numerical methods
will be given in the next section below.

\newsection{Numerical analysis}

In this section we present the different solutions to the dispersion relations
of eqs.~\eqn{dis1}, \eqn{2in} and \eqn{3in} for the scalar $K\pi$, $K\eta$
and $K\eta'$ form factors. The required partial wave amplitudes are taken
from our previous work \cite{JOP:00a}. First, we investigate the single channel
elastic case for which the analytical Omn\`es solution is available. Then the
inelastic two-- and three--channel problems are treated for which we have to
resort to numerical methods.

\subsection{Elastic $K\pi$ channel}

In the elastic case, the analytical Omn\`es solution is given by eq.~\eqn{omel}.
It only depends on the elastic $K\pi$ phase shift $\delta_1(s)$. However, we
still have to fix the polynomial ambiguity $P(s)$. This can be achieved from
the assumption that $F_1(s)$ should vanish for $s$ going to infinity.

Investigating $F_1(s)$ in this limit, from eq.~\eqn{omel} one finds:
\begin{equation}
\label{limom}
\lim_{s\rightarrow \infty}F_1(s) \; = \; e^{i\delta_{1\infty}}
\lim_{s\rightarrow \infty}P(s)\biggl(\frac{s_{th\,1}}{s}\!\biggr)^
{\delta_{1\infty}/\pi} \,,
\end{equation}
with $\delta_{1\infty}\equiv \lim_{s\rightarrow \infty}\delta_1(s)$. In the
single channel case with one resonance, $\delta_{1\infty}$ is expected to be
equal to $\pi$ \cite{MS:70}. Assuming $F_1(s)$ to vanish at infinity thus
requires that $P(s)$ should be constant. This constant can be determined from
the normalisation of $F_1(s)$ at $s=0$. Then the Omn\`es formula takes the
form:
\begin{equation}
\label{omeln}
F_1(s) \, = \, F_1(0) \exp \left( \frac{s}{\pi} \int\limits^\infty_{s_{th\,1}}
\!\!ds'\, \frac{\delta_1(s')}{s'(s'-s-i0)}\right) \,. 
\end{equation}
In our numerical analysis, for $F_1(0)$ we will use the next-to-leading order
$\chi$PT result of eq.~\eqn{ffxpt}, namely $F_1^{\chi\mbox{\tiny PT}}(0) =
f_{K\pi}(0) = 0.981$. This is very close to the leading order value 1,
demonstrating that at $s=0$, as expected, $\chi$PT works extremely well.

\begin{figure}[htb]
\centerline{
\rotate[r]{
\epsfysize=14cm\epsffile{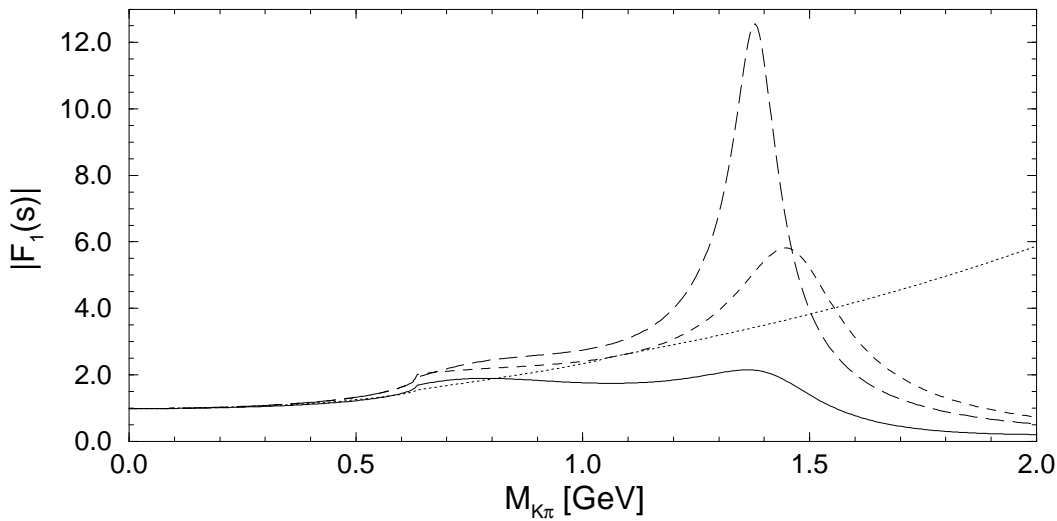} } }
\vspace{-4mm}
\caption[]{$|F_1(s)|$ in the elastic case. Long--dashed line: $\chi$PT fit of
eq.~(4.9) of \cite{JOP:00a}; solid line: $\chi$PT fit of eq.~(4.10) of \cite{
JOP:00a}; short--dashed line: K-matrix fit of eq.~(4.13) of \cite{JOP:00a};
dotted line: next-to-leading order $\chi$PT result of eq.~\eqn{ffxpt}.
\label{fig1}}
\end{figure}

Inputs for the phase shift $\delta_1(s)$, required in eq.~\eqn{omeln},
will be taken from different fits of our previous work on S--wave $K\pi$
scattering \cite{JOP:00a}. In fig.~\ref{fig1} the absolute value of the
form factor, $|F_1(s)|$, is displayed for three different elastic fits, as
a function of the invariant mass of the $K\pi$ system $M_{K\pi}=\sqrt{s}$.
The long--dashed line corresponds to the $\chi$PT fit of eq.~(4.9) of
\cite{JOP:00a}, the solid line to the $\chi$PT fit of eq.~(4.10) of
\cite{JOP:00a}, and finally the short--dashed line corresponds to the
K-matrix fit of eq.~(4.13) of \cite{JOP:00a}. For comparison, as the dotted
line we also show the next-to-leading order $\chi$PT result of eq.~\eqn{ffxpt}.
This comparison shows that up to the $K\pi$ threshold region, next-to-leading
order $\chi$PT gives a good description of the scalar form factor.

As is obvious from fig.~\ref{fig1}, although all three fits give very good
representations of the $K\pi$ scattering data in the elastic region, the
corresponding form factors, especially in the resonance region, behave very
differently. Let us discuss these findings in more detail. In the resonance
region the form factor corresponding to our best elastic fit of eq.~(4.10) of
\cite{JOP:00a} is almost flat, displaying no significant resonance structure.
This can be traced back to the fact that for large $s$, $\delta_1^{(4.10)}(s)$
tends to zero, whereas for the passage of one resonance the phase shift should
go to $\pi$. The $\chi$PT fit of eq.~(4.9) of \cite{JOP:00a}, including the
short--distance constraint discussed in section~2, on the other hand, displays
this behaviour. However, in this case the scalar form factor in the resonance
region is rather large. This is due to the fact that for this fit the width
of the $K_0^*(1430)$ comes out to be very small, about a factor of three below
the experimental average. (See section~7 of \cite{JOP:00a}.) Nevertheless, one
has to strongly stress that a proper study of the resonance region can only
be performed after including the $K\eta'$ with a threshold around $1.45\,\gev$,
as shown below.

With respect to the general behaviour discussed above, the form factor
corresponding to the K-matrix fit of eq.~(4.13) of \cite{JOP:00a} appears to
be the most realistic one. $\delta_1^{(4.13)}(s)$ tends to $\pi$ for large
$s$ and also the resonance parameters of the $K_0^*(1430)$ for this fit are
close to the experimental average. The K-matrix fit is also different from
the chiral fits because the corresponding ansatz for the partial wave amplitude
is free of left--hand cuts. In this case it is even possible to give a closed
expression for the form factor \cite{babelon,MS:70}:
\begin{equation}
\label{ffKmat}
F_1(s) \;=\; \prod\limits_{i,j} \frac{(1-s/s_{p_i})}{(1-s/s_{z_j})}\,
\frac{F_1(0)}{\Big(1-C(s)K_0^{1/2}(s)\Big)} \,,
\end{equation}
where $C(s)=16\pi\bar J_{K\pi}(s)$ and the explicit expression for the
K-matrix ansatz $K_0^{1/2}(s)$ is given in eq.~(4.12) of \cite{JOP:00a}.
The $s_{p_i}$ and $s_{z_i}$ are the locations of the poles and zeros of
$(1-C(s)K(s))^{-1}$ which have to be removed in the form factor. In our
particular case we have one pole at $s_{p_1}=-2.011\,\gev^2$ and two zeros at
$s_{z_1}=-5.821\,\gev^2$ and $s_{z_2}=M_{K_0^*(1430)}^2$. It is an easy matter
to verify that the representation \eqn{ffKmat} yields the same scalar form
factor as the Omn\`es formula \eqn{omeln}.

\subsection{The two--channel case}

We now turn to the two--channel case, for which the important effects
of the $K\eta'$ are considered in the coupled channel integral equations
\eqn{2in}. Let us first briefly describe our numerical treatment of these
integral equations.

The first step consists in rewriting the coupled integral equations according
to the procedure described in appendix~C.1. As central values for the two 
parameters which are introduced in this context we take $s_{cut} = 9\;\gev^2$
and $b=-\,0.8$. We have, however, varied these parameters to test the numerical
stability of our approach.\footnote{One of us, J.A.O., has implemented a
different numerical approach, finding agreement for the resulting form factors.}
Next, the resulting integral equations are solved iteratively according to a
procedure already employed in ref.~\cite{dona:90}. 

For the iterative solution we require initial functions $F_1^{(0)}(s)$ and
$F_3^{(0)}(s)$. For $F_1^{(0)}(s)$, we use the form factor which results from
employing the Omn\`es formula \eqn{omeln}. The phase-shift $\delta_1(s)$ is
taken according to the decomposition \eqn{S}. As a starting point, the form
factor $F_3^{(0)}(s)$ can be set to zero. We have verified that other initial
choices for this function lead to the same final form factors.

Since the dispersion integrals are calculated according to the principal value
prescription, the immediate results of our first iteration step are the real
parts $\RE F_1^{(1)}(s)$ and $\RE F_3^{(1)}(s)$. From these results, together
with the initial imaginary parts $\IM F_1^{(0)}$ and $\IM F_3^{(0)}$, and
making use of the central unitarity relation \eqn{unif}, the new full form
factors $F_1^{(1)}$ and $F_3^{(1)}$ are obtained. This procedure is repeated,
until after $n$ steps it has converged reasonably well and the final form
factors $F_1^{(n)}$ and $F_3^{(n)}$ can be extracted.

In what follows, we shall explicitly investigate the form factors corresponding
to the two--channel fits (6.10) and (6.11) of ref.~\cite{JOP:00a} to the S--wave
$K\pi$-scattering data. In a forthcoming publication \cite{JOP:01}, these form
factors and the corresponding scalar spectral function will be utilised to
calculate the mass of the strange quark in the framework of QCD sum rules.

\begin{figure}[bht]
\centerline{
\rotate[r]{
\epsfysize=14cm\epsffile{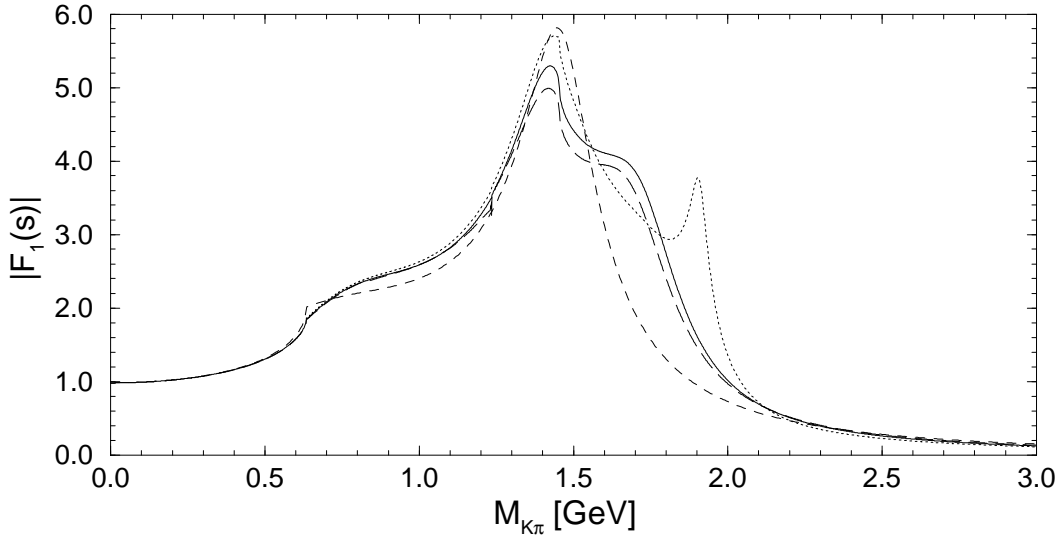} } }
\vspace{-4mm}
\caption[]{$|F_1(s)|$ in the two--channel case. The solid and long--dashed lines
correspond to the $\chi$PT fits of eqs.~(6.10) and (6.11) of \cite{JOP:00a}
respectively. For comparison, the elastic K-matrix fit of eq.~(4.13) is
displayed as the short--dashed line, and the dotted line represents the Omn\`es
formula \eqn{omeln}, evaluated with $\delta_1^{(6.10)}(s)$.

\label{fig2}}
\end{figure}
\begin{figure}[thb]
\centerline{
\rotate[r]{
\epsfysize=14cm\epsffile{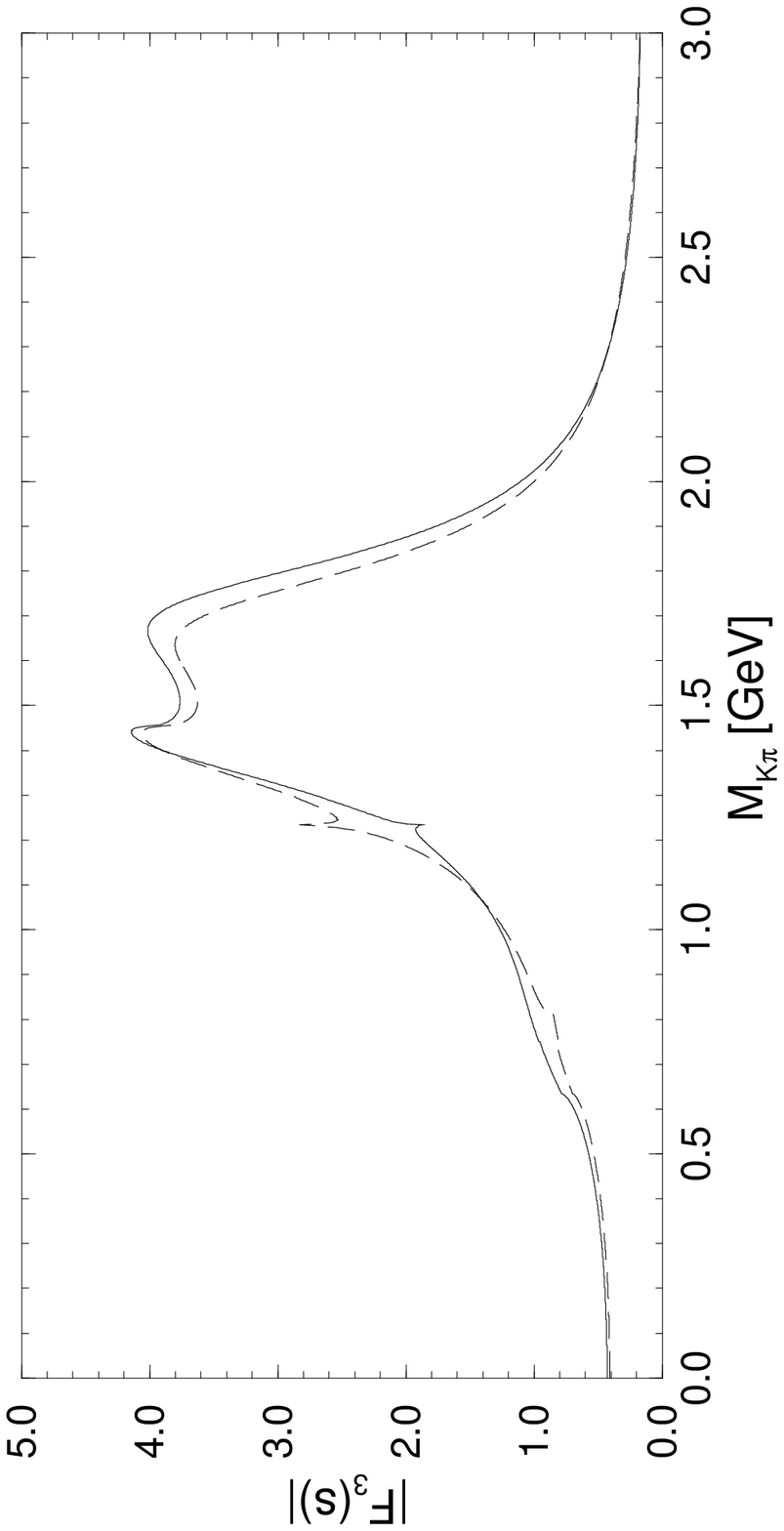} } }
\vspace{-4mm}
\caption[]{$|F_3(s)|$ in the two--channel case. The solid and long--dashed lines
correspond to the $\chi$PT fits of eqs.~(6.10) and (6.11) of \cite{JOP:00a}
respectively.\label{fig3}}
\end{figure}

The absolute value of the resulting inelastic two--channel $K\pi$ and $K\eta'$
form factors $F_1(s)$ and $F_3(s)$ is displayed in figures~\ref{fig2} and
\ref{fig3}. The solid line corresponds to the $\chi$PT fit of eq.~(6.10) of
ref.~\cite{JOP:00a} and the long--dashed line to the fit (6.11). Depending on
the high-energy behaviour of the phase shifts $\delta_1(s)$ and $\delta_3(s)$,
the dispersion relations admit one or two linearly independent solutions
\cite{babelon}. For the fits (6.10) and (6.11), with $s$ going to infinity,
$\delta_1(s)$ tends to $\pi$ whereas $\delta_3(s)$ tends to zero. In this case
there is only one solution to the two--channel dispersion relation. However,
we still have the freedom to normalise both solutions $F_1(s)$ and $F_3(s)$ by
a common factor. We have used this freedom to again fix $F_1^{\chi\mbox{\tiny
PT}}(0) = f_{K\pi}(0) = 0.981$.

The value of the $K\eta'$ form factor at zero momentum is then determined and
we obtain $F_3^{(6.10)}(0) = 0.430$ and $F_3^{(6.11)}(0) = 0.409$. These
results should be compared with the tree--level expectation from ${\rm U}(3)_L
\!\times\!{\rm U}(3)_R$ $\chi$PT in the large-$N_c$ limit including resonances
of eq.~\eqn{F3ChPT}, $F_3^{\chi\mbox{\tiny PT}}(0) = 2\sqrt{2}/3 f_{K\eta'}(0)=
0.696$. We observe that the resulting value for $F_3(0)$ from the solution of
the coupled channel dispersion relation turns out to be somewhat lower than the
tree--level result. We shall come back to a discussion of this point below.

For comparison, as the dotted line in figure~\ref{fig2}, we have displayed the
form factor that results from evaluating the Omn\`es formula eq.~\eqn{omeln}
with $\delta_1^{(6.10)}(s)$. $F_1^{\mbox{\tiny Omn\`es}}(s)$ has been used as
the starting point for the iteration of the integral equations. From figure
\ref{fig2} it is clear that, apart from the region of the second resonance, the
Omn\`es representation of $F_1(s)$ already gives a reasonable description of
this form factor. In addition, as the short--dashed line in figure \ref{fig2}
we have also plotted $|F_1(s)|$ corresponding to the elastic K-matrix fit of
eq.~(4.13) of \cite{JOP:00a}. Here we observe that the corresponding form
factor  behaves rather differently in the $K\pi$ threshold region. This should
come as no surprise, because as has been already discussed at the end of
section~4 of ref.~\cite{JOP:00a}, also the scattering lengths for this fit
come out very different compared to $\chi$PT.

As was already mentioned above, for the fits $(6.10)$ and $(6.11)$ the total
phase motion of $\delta_1(s)+\delta_3(s)$ for $s$ going to infinity only
reaches $\pi$, whereas on general grounds for two resonances it is expected to
approach $2\pi$. The reason for this behaviour of our unitarised chiral fits
can be traced back to a deficient description of the experimental data
\cite{ast:88} above roughly $1.9\;\gev$. To improve the description in the
region of the second resonance and above, we have performed new fits with a
K-matrix ansatz. In the fitting process, the K-matrix ansatz is matched
smoothly to the chiral fits at an energy around $1.75\;\gev$.

\begin{figure}[htb]
\centerline{
\rotate[r]{
\epsfysize=15cm\epsffile{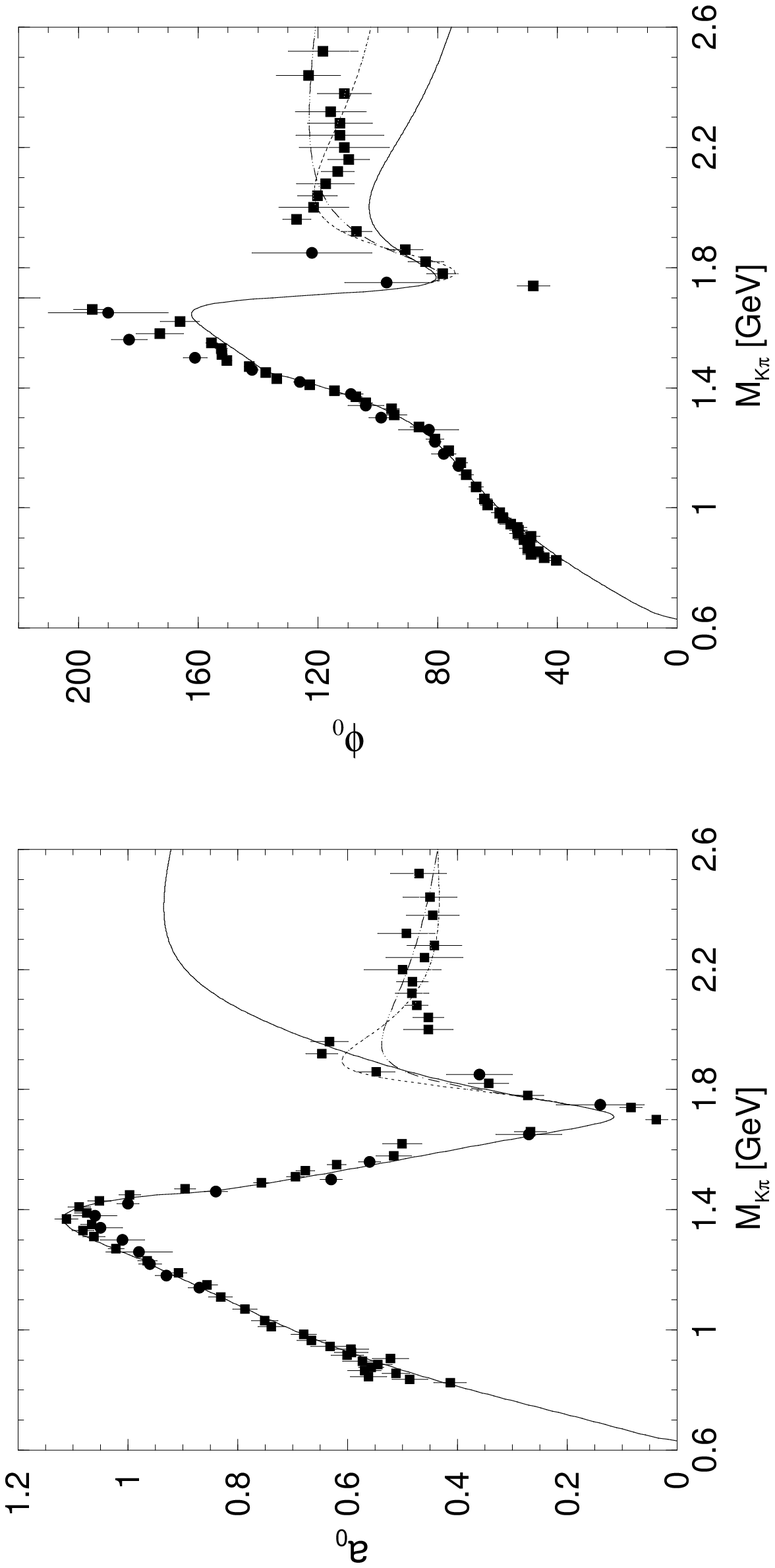} } }
\vspace{-4mm}
\caption[]{Two-channel fits for $a_0$ and $\phi_0$. The experimental data are
given by: solution B of \cite{est:78} full circles; solution A of \cite{ast:88}
full squares. The solid line represents our unitarised chiral fit (6.10) of
ref. \cite{JOP:00a}, and, as examples, the dashed-dotted and dotted lines show
the new K-matrix fits (6.10K1) and (6.10K4) respectively, which display an
improved behaviour in the region of the second resonance.
\label{figkpi}}
\end{figure}

For the K-matrix we use a resonance plus background parametrisation:
\begin{equation}
\label{kmathigh}
K_{ij}(s) \;=\; \frac{r_i r_j}{(M_{S'}^2-s)} +
                \frac{(a_{ij}+b_{ij}s)}{[1+(s/c)^\kappa]} \,,
\end{equation}
with $\kappa=0,\,1$ if $b_{ij}=0$ and $\kappa=2$ otherwise. Except for the
first case $\kappa=0$, this K-matrix is constructed such that it vanishes
linearly at large $s$, and above $1.75\;\gev$ it replaces the matrix $N(s)$ in
eq.~(6.1) of ref.~\cite{JOP:00a}. Because of time-reversal invariance, $K(s)$
has to be symmetric. Above $1.75\;\gev$, we then fit this ansatz to the data
set A of \cite{ast:88} which extends up to $2.52\;\gev$. Simultaneously a
smooth matching to either one of the unitarised chiral fits (6.10) or (6.11)
of \cite{JOP:00a} at around $1.75\;\gev$ is imposed in the fitting procedure
as well. To judge the dependence of the K-matrix fits on the background
parametrisation, we have actually calculated four different types of fits. All
fits have been performed with the program Minuit \cite{min:94}, and the
parameters of these fits are compiled in table~\ref{tab1}. As examples, in
figure~\ref{figkpi} we display the experimental $K\pi$ scattering data of
refs.~\cite{est:78,ast:88}, together with the unitarised chiral fit (6.10) of
\cite{JOP:00a} (solid line) and the new K-matrix fits with improved behaviour
in the region of the second resonance (6.10K1) (dashed-dotted line) and (6.10K4)
(dotted line). The corresponding curves for the other fits look very similar.

\begin{table}[htb]
\begin{center}
\begin{tabular}{crrrrrrrr}
\hline
Fit    & \fz{(6.10K1)} & \fz{(6.10K2)} & \fz{(6.10K3)} & \fz{(6.10K4)} 
       & \fz{(6.11K1)} & \fz{(6.11K2)} & \fz{(6.11K3)} & \fz{(6.11K4)} \\ 
\hline
$M_S'$ [\fgev]            &  $2.040$ &  $2.040$ &  $1.904$ &  $1.909$
                          &  $1.837$ &  $1.837$ &  $1.838$ &  $1.837$ \\
$r_1$ [\fgev]             &  $1.007$ &  $1.003$ &  $0.567$ &  $0.617$
                          &  $0.495$ &  $0.501$ &  $0.449$ &  $0.433$ \\
$r_2$ [\fgev]             &  $1.420$ &  $1.406$ &  $0.550$ &  $0.648$
                          &  $0.595$ &  $0.603$ &  $0.499$ &  $0.474$ \\
$a_{11}$                  & $-0.723$ & $-0.801$ & $-0.703$ &  $0.221$
                          & $-0.350$ & $-0.407$ & $-0.647$ & $-0.346$ \\
$a_{12}$                  & $-0.905$ & $-0.990$ &  $0.495$ &  $1.797$
                          & $-0.356$ & $-0.415$ &  $0.565$ &  $1.741$ \\
$a_{22}$                  & $-1.450$ & $-1.580$ & $-1.276$ & $-0.159$
                          & $-0.567$ & $-0.656$ & $-0.390$ &  $0.750$ \\
$b_{11}$ [\fgev${}^{-2}$] &      $0$ &      $0$ &  $0.106$ & $-0.308$
                          &      $0$ &      $0$ &  $0.145$ &  $0.028$ \\
$b_{12}$ [\fgev${}^{-2}$] &      $0$ &      $0$ & $-0.214$ & $-0.738$
                          &      $0$ &      $0$ & $-0.224$ & $-0.606$ \\
$b_{22}$ [\fgev${}^{-2}$] &      $0$ &      $0$ &  $0.366$ & $-0.123$
                          &      $0$ &      $0$ &  $0.055$ & $-0.326$ \\
$c$ [\fgev${}^2$]         & $\infty$ &     $25$ &     $25$ &      $4$
                          & $\infty$ &     $25$ &     $25$ &      $4$ \\
$\chi^2/33\;$dof          &   $74.9$ &   $75.5$ &   $42.3$ &   $45.2$
                          &   $83.3$ &   $87.4$ &   $43.2$ &   $38.0$ \\
$\delta_{1\infty},\,\delta_{3\infty}$ & $\pi,\,0$ & $\pi,\,\pi$ & $2\pi,\,0$ &
$2\pi,\,0$ & $\pi,\,0$ & $\pi,\,\pi$ & $\pi,\,\pi$ & $2\pi,\,0$ \\
\hline
\end{tabular}
\end{center}
\caption{Different K-matrix fits corresponding to the fits (6.10) and (6.11)
of \cite{JOP:00a} in the region above $1.75\;\gev$. For a detailed discussion
see the text.
\label{tab1}} 
\end{table}

For the first type of fits (6.10K1) and (6.11K1), as the background we have
just taken constants $a_{ij}$. This entails that the T-matrix at large $s$
does only fall off as $1/\ln(s)$, and that the total phase motion at infinity,
$\delta_{1\infty}+\delta_{3\infty}$ is only $\pi$, analogous to the case of the
unitarised chiral fits. For the convenience of the reader, the separate phases
at infinity, $\delta_{1\infty}$ and $\delta_{3\infty}$, have also been collected
in table~\ref{tab1}. For these two fits, again $F_3(0)$ is determined, and
comes out as $F_3^{(6.10K1)}(0) = 0.591$ and $F_3^{(6.11K1)}(0) = 0.574$ for
(6.10K1) and (6.11K1) respectively, much closer to the tree-level result
$F_3^{\chi\mbox{\tiny PT}}(0) = 0.696$. As was already remarked above, at the
tree--level the scalar resonance mass $M_S$ has some ambiguity, and e.g. using
$M_S=1.2\;\gev$ in eq.~\eqn{F3ChPT} would yield $F_3^{\chi\mbox{\tiny PT}}(0)=
0.586$, in perfect agreement with the fit results for $F_3(0)$.

To further investigate the quality of our fits, in the following we shall
include additional low-energy constraints on the scalar $K\pi$ form factor
which come from pure next-to-leading order $\chi$PT \cite{GL:85}. Rather
precise information on $F_1(s)$ is available at the so called Callan-Treiman
point $M_K^2-M_\pi^2$, because at that point $F_1(\DeKP)$ only differs from
$F_K/F_\pi$ by terms of order $m_u$ or $m_d$,
\begin{equation}
\label{CTrel}
F_1(\DeKP) \;=\; \frac{F_K}{F_\pi} + \Delta_{CT} \,,
\end{equation}
where numerically $\Delta_{CT}=-3.0\cdot 10^{-3}$ at the next-to-leading order
in $\chi$PT, and the analytic expression can be found in the second of refs.
\cite{GL:85}. Taking most recent values for the decay constants $F_K$ and
$F_\pi$ from \cite{pdg:00}, we conclude that $F_1(\DeKP) = 1.22 \pm 0.01$.

Furthermore, from $\chi$PT we also have information on the slope of the $K\pi$
form factor at zero energy, $D_1(0)\equiv F_1'(0)/F_1(0)$. In $\chi$PT at the
next-to-leading order, one obtains $D_1(0)=0.873\;\gev^{-2}$, where the
uncertainty due to higher order corrections was estimated to be around 20\%
\cite{GL:85}. This value can also be compared to experimental data. The slope
of the scalar form factor has been measured in charged as well as neutral Kaon
decays, with the results $\lambda_0 = 0.006\pm 0.007$ and $\lambda_0 = 0.025\pm
0.006$ respectively \cite{pdg:00}. Here, the parameter $\lambda_0$ is defined
by $\lambda_0 \equiv D_1(0) M_\pi^2$. Taking the average of the experimental
results and enlarging the error according to the standard procedure of the
Particle Data Group for inconsistent measurements, we obtain $\lambda_0 =
0.017\pm 0.009$, whereas from the $\chi$PT result together with the isospin
average $M_\pi = 138\;\mev$, we also find $\lambda_0 = 0.017$, displaying
complete agreement for the central results.

Let us now compare $D_1(0)$ and especially $F_1(\DeKP)$ with the numbers
emerging from our fits. For the K-matrix fits which accomplish a good
description of the high energy data above $1.75\;\gev$, we get $F_1^{(6.10K1)}
(\DeKP)=1.220$ and $F_1^{(6.11K1)}(\DeKP)=1.219$, in impressive agreement with
the precise $\chi$PT expectation of eq.~\eqn{CTrel}. In addition, the
slope is found to be $D_1^{(6.10K1)}(0)=0.894\;\gev^{-2}$ as well as
$D_1^{(6.11K1)}(0)=0.893\;\gev^{-2}$, also very close to the $\chi$PT value.
On the other hand, for the original unitarised chiral fits the corresponding
results are $F_1^{(6.10)}(\DeKP)=1.262$ and $F_1^{(6.11)} (\DeKP)=1.260$, as
well as $D_1^{(6.10)}(0)=0.918\;\gev^{-2}$ and $D_1^{(6.11)}(0)=0.909\;
\gev^{-2}$. Although the values for $D_1(0)$ are compatible with our
previous results, $F_1(\DeKP)$ turns out too large for both fits, again
reflecting an insufficient description of the higher energy region for
these fits.

\begin{table}[htb]
\begin{center}
\begin{tabular}{ccc|ccc}
\hline
$F_1(\DeKP)$ & $F_3(0)$ & $D_1(0)$ [\fgev${}^{-2}$] &
$F_1(\DeKP)$ & $F_3(0)$ & $D_1(0)$ [\fgev${}^{-2}$] \\
\hline
& (6.10K2) & & & (6.11K2) & \\
1.21 & 0.636 & 0.848 & 1.21 & 0.612 & 0.850 \\
1.22 & 0.587 & 0.863 & 1.22 & 0.565 & 0.866 \\
1.23 & 0.538 & 0.879 & 1.23 & 0.518 & 0.882 \\
\hline
& (6.10K3) & & & (6.11K3) & \\
1.21 & 0.586 & 0.828 & 1.21 & 0.618 & 0.777 \\
1.22 & 0.567 & 0.851 & 1.22 & 0.587 & 0.805 \\
1.23 & 0.547 & 0.874 & 1.23 & 0.556 & 0.834 \\
\hline
& (6.10K4) & & & (6.11K4) & \\
1.21 & 0.622 & 0.851 & 1.21 & 0.615 & 0.858 \\
1.22 & 0.590 & 0.868 & 1.22 & 0.579 & 0.874 \\
1.23 & 0.557 & 0.886 & 1.23 & 0.544 & 0.890 \\
\hline
\end{tabular}
\end{center}
\caption{$F_3(0)$ and $D_1(0)$ for the unitarised chiral plus K-matrix fits
(6.10K2--4) and (6.11K2--4) with $\delta_{1\infty}+\delta_{3\infty}=2\pi$,
chosen such that $F_1(\DeKP)=1.22\pm 0.01$.
\label{tab2}}
\end{table}

For the previous fits, the total phase motion $\delta_{1\infty}+
\delta_{3\infty}$ was only $\pi$, although for two resonances $2\pi$ would be
expected. The differences are due to the background accompanying the bare
K-matrix poles in eq.~\eqn{kmathigh}. The latter behaviour can be achieved by
forcing the K-matrix to vanish at infinity as will be implemented in all our
remaining fits. In this way, we also test the stability of our results under
changes in the parameterisations that induce different T-matrices well above
the second resonance region around $2\;\gev$. As a first step, in the fits
(6.10K2) and (6.11K2), we introduce an additional cutoff parameter $c=25\;
\gev^2$. As long as $c$ is large enough, the sensitivity on this parameter is
rather small and it cannot be incorporated as a fit parameter. As can be seen
from table \ref{tab1}, the remaining fit parameters do depend only little on
this modification. Nevertheless, now both phase shifts $\delta_1$ and $\delta_3$
approach $\pi$ at infinity. As a consequence, the dispersion relations \eqn{2in}
admit two linearly independent solutions. Thus, besides $F_1(0)$, now we are in
a position to also fix $F_3(0)$ as an initial condition. In our opinion the
best information available is $F_1(\DeKP)$, and therefore we fix $F_3(0)$ such
as to fullfill the constraint on this quantity presented above.

Our results for this exercise are shown in table~\ref{tab2}. Before we come
to a detailed discussion of these results, however, let us first present the
remaining fits of table~\ref{tab1}. As can be read off this table, the
$\chi^2$ of the K-matrix fits (6.10K1), (6.10K2), (6.11K1) and (6.11K2) is
larger than 2 for all cases. Let us remark that this $\chi^2$ only corresponds
to the data points of \cite{ast:88} above $1.85\;\gev$, which have not yet
been included in the fits (6.10) and (6.11), and the matching condition around
$1.75\;\gev$. To improve the quality of our K-matrix fits, we have decided to
include an additional term $b_{ij}s$ in the background. To maintain the required
falloff of the T-matrix, for these fits the parameter $\kappa=2$. In order to
study the cutoff dependence we have performed this type of fits with the two
values $c=25\;\gev^2$ and a lower cutoff $c=4\;\gev^2$.

For the fits (6.10K3), (6.10K4) and (6.11K4), we observe that the phase shifts
at infinity turn out to be $\delta_{1\infty}=2\pi$ and $\delta_{3\infty}=0$.
The reason for this change in the behaviour of the phase shifts lies in the
fact that the second resonance couples almost equally to the first and the
third channel, so that a small change in the fit parameters can shift the
phase motion of $\pi$ from the third to the first channel. For the parameter
set where this happens, we also observe a vanishing inelasticity and thus
zeros in the diagonal elements of the S-matrix. Therefore the phases
$\delta_1(s)$ and $\delta_3(s)$ at this point become ambiguous, although their
sum must be continuous (modulo $2\pi$) since the matrix element $S_{12}$ of
eq.~\eqn{S} is non-vanishing. As far as the phase shifts at infinity are
concerned, the fit (6.11K3) behaves as (6.10K2) and (6.11K2). However, this fit
displays an artificial rapid phase motion at an energy around $13\;\gev$, and
although it has a rather good $\chi^2$ and the results for $F_3(0)$ in table
\ref{tab2} turn out reasonable, we shall discard it in the following.

\begin{figure}[htb]
\centerline{
\rotate[r]{
\epsfysize=14cm\epsffile{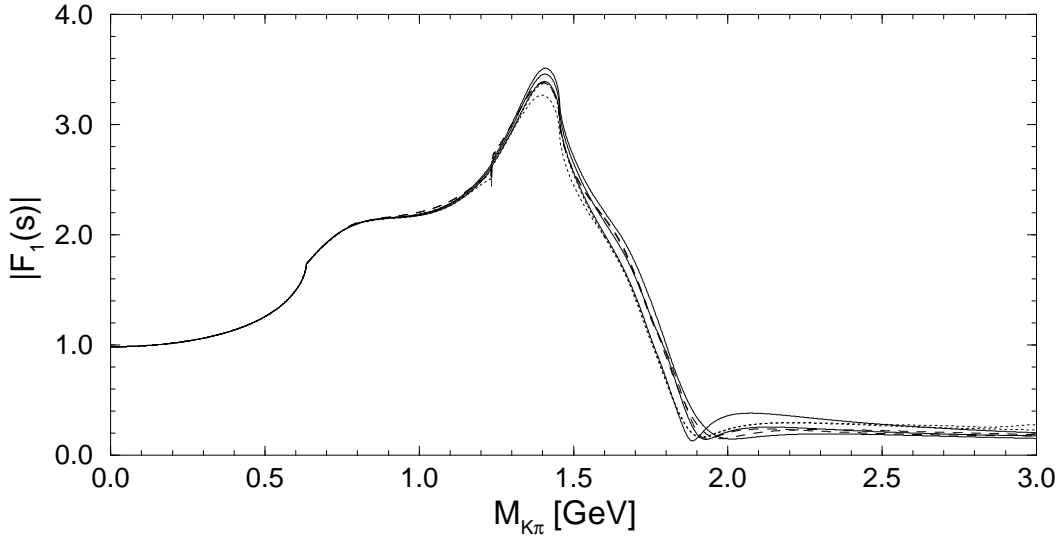} } }
\vspace{-4mm}
\caption[]{$|F_1(s)|$ in the two--channel case with K-matrix ansatz. The
dotted lines correspond to the fits (6.10K1) and (6.11K1), the solid lines to
(6.10K2), (6.10K3) and (6.10K4), and the dashed lines to (6.11K2) and (6.11K4).
\label{fig4}}
\end{figure}

\begin{figure}[htb]
\centerline{
\rotate[r]{
\epsfysize=14cm\epsffile{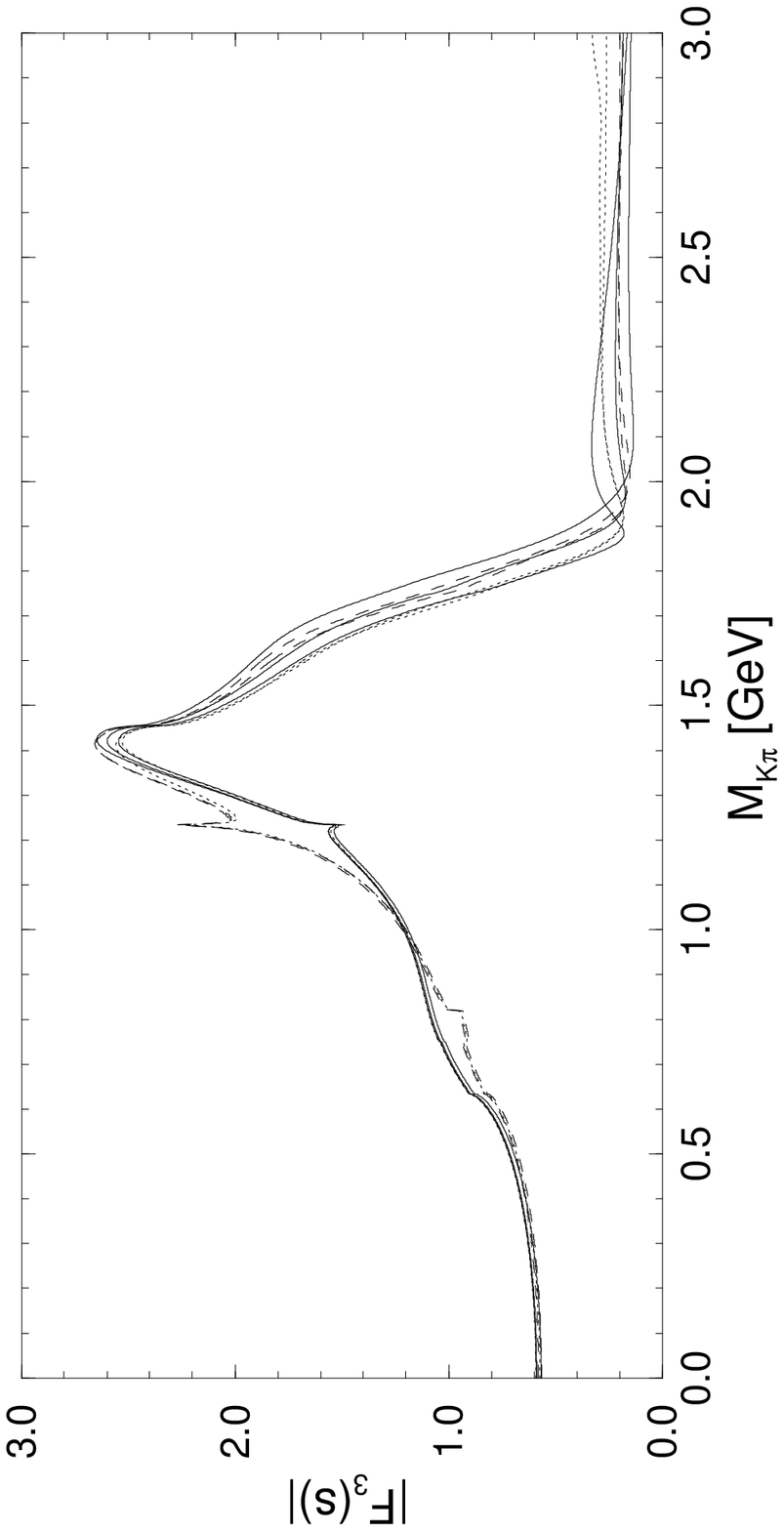} } }
\vspace{-4mm}
\caption[]{$|F_3(s)|$ in the two--channel case with K-matrix ansatz. The
dotted lines correspond to the fits (6.10K1) and (6.11K1), the solid lines to
(6.10K2), (6.10K3) and (6.10K4), and the dashed lines to (6.11K2) and (6.11K4).
\label{fig5}}
\end{figure}

Our results for the absolute values of the form factors $F_1(s)$ and $F_3(s)$
in the case of the unitarised chiral plus K-matrix fits are shown in figures
\ref{fig4} and \ref{fig5}. The dotted lines represent the fits (6.10K1) and
(6.11K1), the solid lines (6.10K2), (6.10K3) and (6.10K4), and finally, the
dashed lines correspond to (6.11K2) and (6.11K4). Since all form factors for
these fits are rather similar we have not tried to discriminate the curves
further. Nevertheless, for $|F_3(s)|$, like in figure \ref{fig3}, there are
differences in the form factor between the fit (6.10) and (6.11) which are most
prominent around $1.2\;\gev$. These differences can already be seen in the
T-matrix and they tend us to consider the fit (6.10) more physical, as was
already concluded in ref.~\cite{JOP:00a}.

\begin{figure}[htb]
\centerline{
\rotate[r]{
\epsfysize=14cm\epsffile{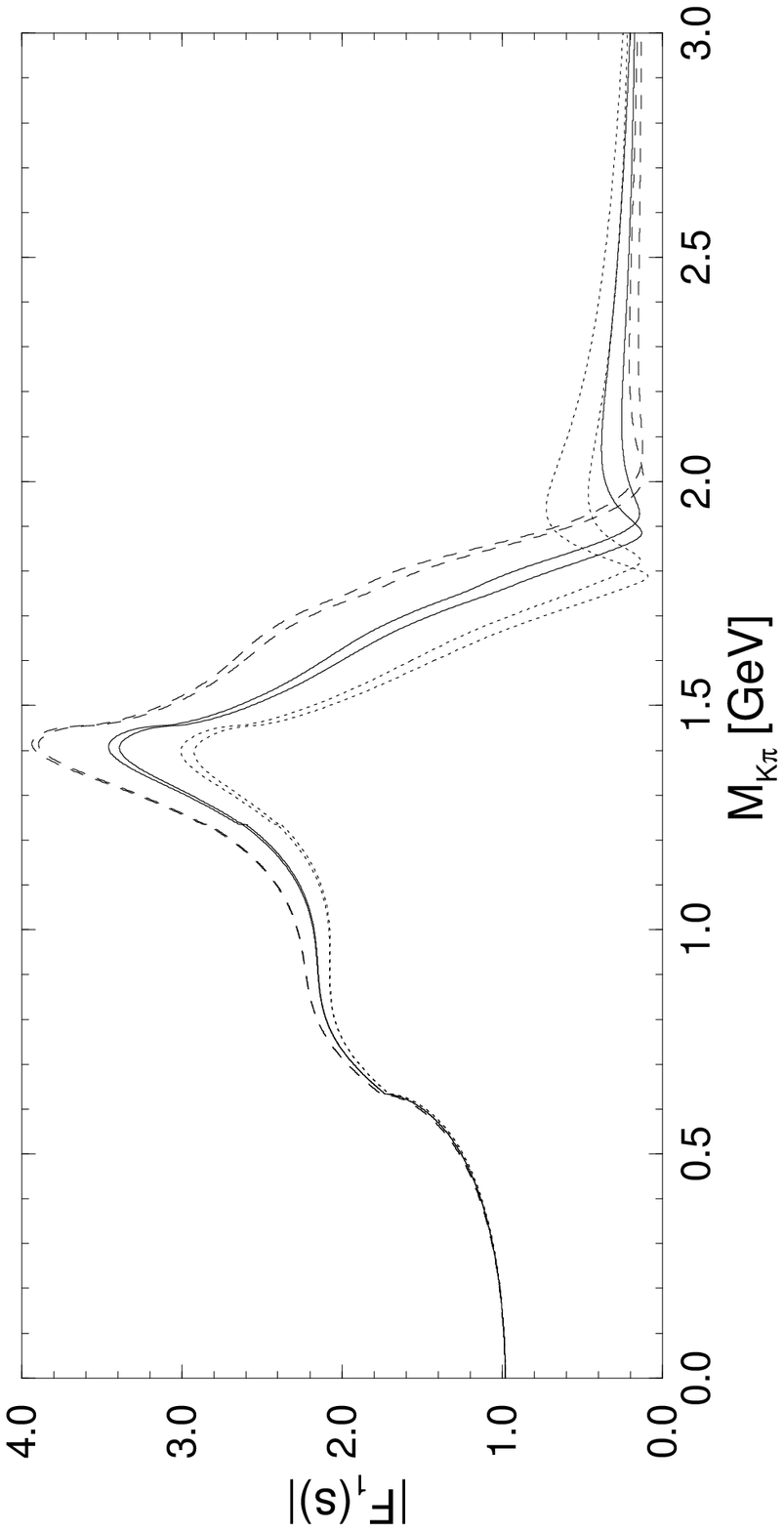} } }
\vspace{-4mm}
\caption[]{$|F_1(s)|$ for the fits (6.10K3) and (6.10K4) while varying
$F_1(\DeKP)$. The dotted lines correspond to $F_1(\DeKP)=1.21$, the solid lines
to $1.22$, and the dashed lines to $1.23$. The curves which are higher at the
$K_0^*(1430)$ resonance always correspond to (6.10K4).
\label{fig6}}
\end{figure}

\begin{figure}[htb]
\centerline{
\rotate[r]{
\epsfysize=14cm\epsffile{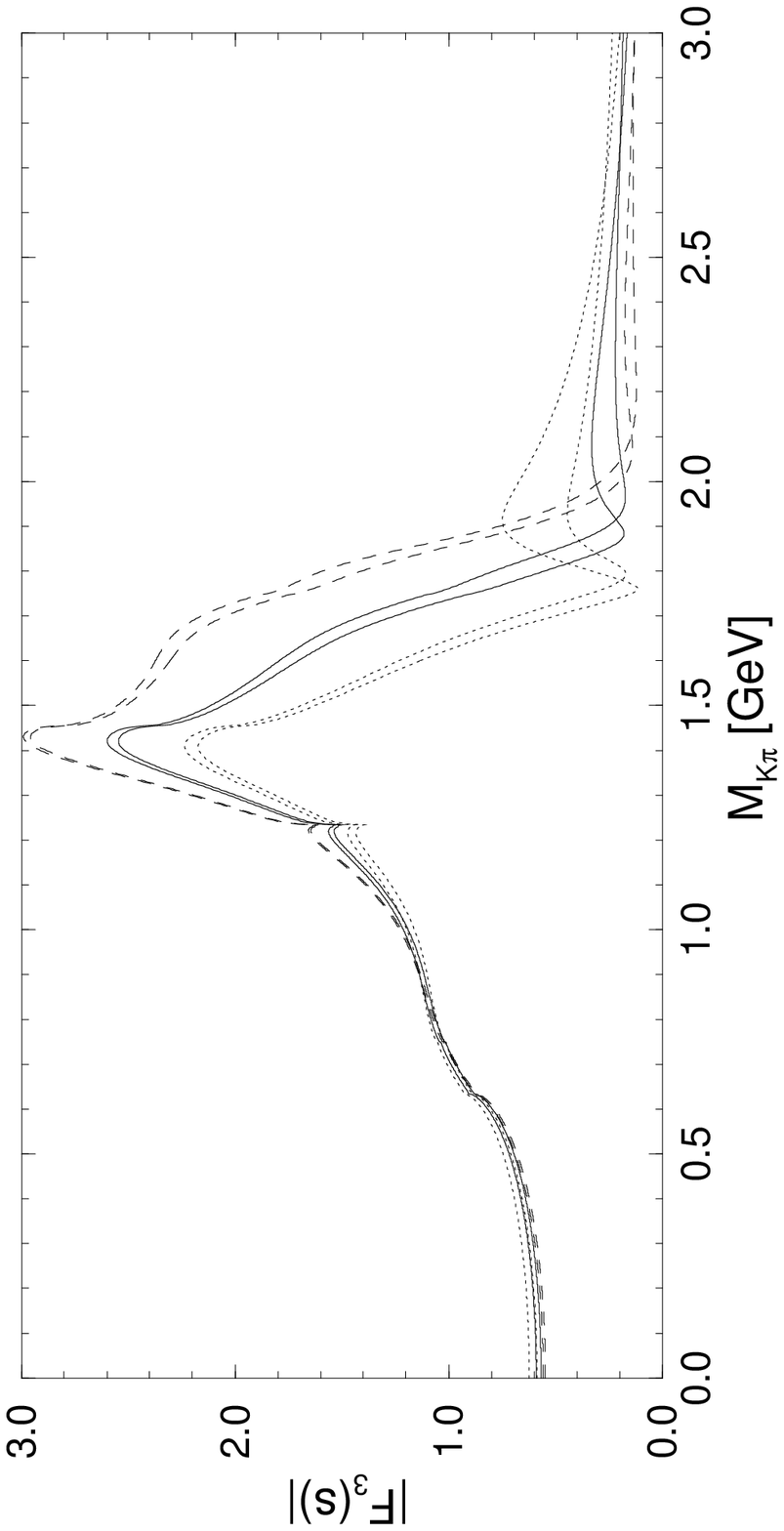} } }
\vspace{-4mm}
\caption[]{$|F_3(s)|$ for the fits (6.10K3) and (6.10K4) while varying
$F_1(\DeKP)$. The dotted lines correspond to $F_1(\DeKP)=1.21$, the solid lines
to $1.22$, and the dashed lines to $1.23$. The curves which are higher at the
$K_0^*(1430)$ resonance always correspond to (6.10K4).
\label{fig7}}
\end{figure}

We shall return now to a discussion of table~\ref{tab2}. In figures~\ref{fig6}
and \ref{fig7}, we display $|F_1(s)|$ and $|F_3(s)|$ for our fits (6.10K3) and
(6.10K4), while varying $F_1(\DeKP)$. The dotted lines correspond to $F_1(\DeKP)
= 1.21$, the solid lines to $1.22$, and the dashed lines to $1.23$. The curves
which are higher at the $K_0^*(1430)$ resonance always represent the fit
(6.10K4). As can be seen from table~\ref{tab2}, for smaller values of
$F_1(\DeKP)$, $F_3(0)$ comes out larger and vice versa. Averaging our results
for the K-matrix fits, for $F_3(0)$ we then extract
\begin{equation}
\label{F30}
F_{K\eta'}(0) \;=\; 0.58 \pm 0.04 \,.
\end{equation}
As the uncertainty we have chosen the maximal variation for the fits K3 and K4
with the better $\chi^2$. The slightly larger variations in the case of K2, to
our minds are due to the worse description of the experimental data. Also the
derivative of the form factor $F_1(s)$ at zero, $D_1(0)$, comes out completely
consistent with our expectation from $\chi$PT, with some variation for the fit
(6.11K3), which anyway has been discarded for other reasons, although even this
variation stays within the uncertainties. Finally, it is worth to note here
that the predicted values for $F_3(0)$ from the fits (6.10K1) and (6.11K1) are
in perfect agreement with the previous central value of $F_3(0)$, and their
difference is completely accounted for by the estimated error in eq.~\eqn{F30}.

\subsection{The three--channel case}

In the following, we investigate the solutions of the dispersion relations
\eqn{3in} in the full three--channel case for our unitarised chiral fits (6.10)
and (6.11) of ref.~\cite{JOP:00a}. In the three--channel case, we have not
improved our fits in the region of the second resonance, because on the one
hand just from the $K\pi$ scattering data there is not enough information to
fit all parameters of a three-channel K-matrix, and, on the other hand, as was
already mentioned above, the effects of the $K\eta$ channel will turn out so
small that the more elaborate two--channel analysis with only $K\pi$ and
$K\eta'$ for our purposes is completely sufficient.\footnote{We have not
taken into account the fit (6.7) of ref.~\cite{JOP:00a}, because when the
$K\eta$ channel is removed for this fit, the $\chi^2$ becomes rather large,
increasing by almost a factor of two. On the theoretical side, this fit does
not fullfill the short distance constraints presented in section~2. The strong
sensitivity of the fit (6.7) on the $K\eta$ channel, and also the very small
value of $c_m$, tend us to disregard this fit for further analysis.} As we
shall verify below, this also entails that the form factors $F_1(s)$ and
$F_3(s)$ are completely stable under the inclusion of the scalar $K\eta$ form
factor $F_2(s)$. Therefore, this form factor will play a negligible role in the
determination of the mass of the strange quark \cite{JOP:01}.

The numerical solution of the dispersion relations proceeds along the same
lines as was already discussed in the last section for the two--channel case.
Again, as initial functions for the form factors we can take the Omn\`es
solution for $F_1^{(0)}(s)$, whereas the form factors $F_2^{(0)}(s)$ and
$F_3^{(0)}(s)$ can be set to zero. Nevertheless, in order to prove the
stability of the two--channel solutions (6.10) and (6.11) under the inclusion
of the $K\eta$ channel, we have also used these solutions as initial functions
$F_1^{(0)}(s)$ and $F_3^{(0)}(s)$ in the present three--channel case. Then the
solution of the dispersion relations are iterated, until the procedure has
converged and the form factors $F_1(s)$, $F_2(s)$ and $F_3(s)$ can be extracted.
Of course, with both choices, the same final form factors are obtained.

\begin{figure}[htb]
\centerline{
\rotate[r]{
\epsfysize=14cm\epsffile{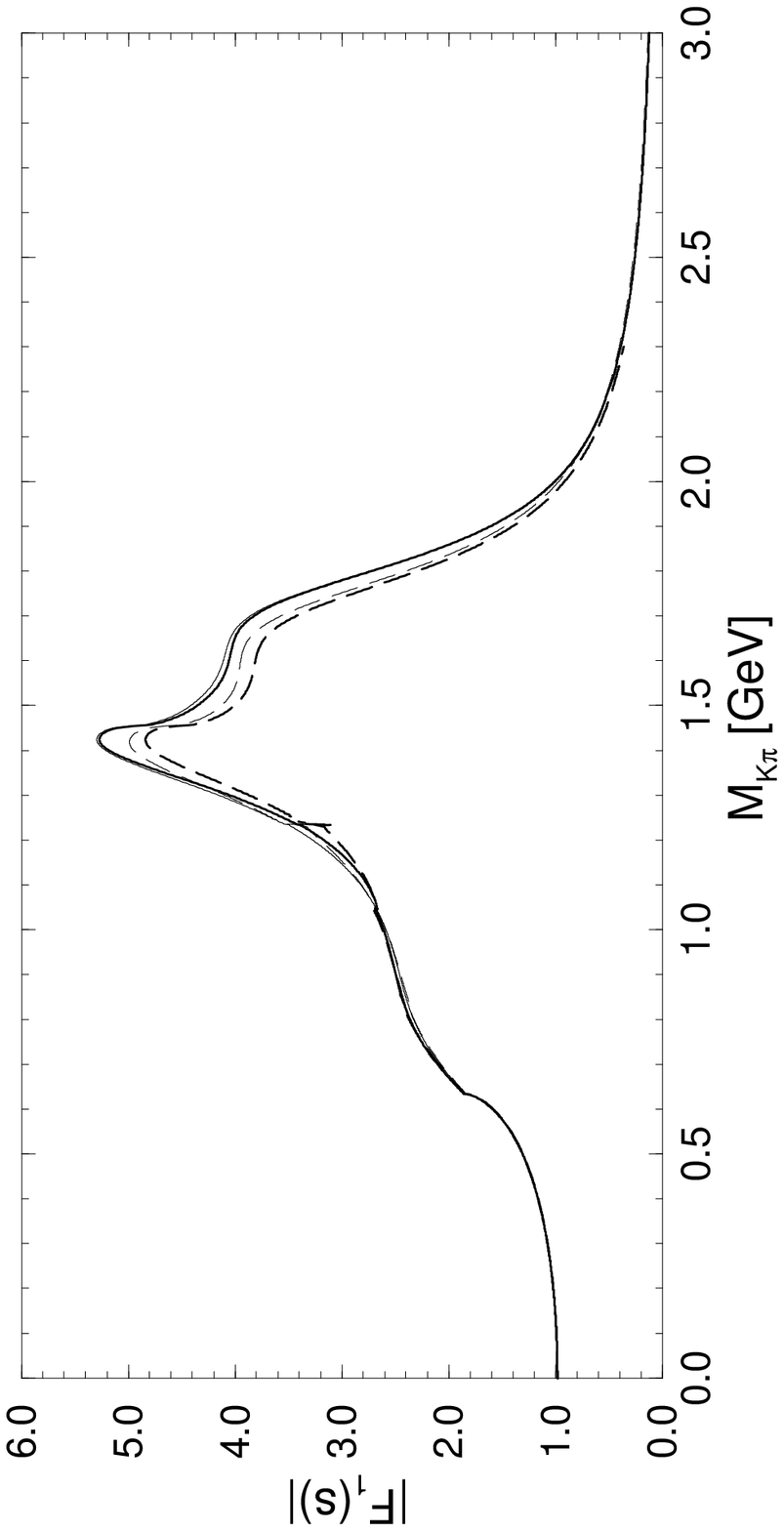} } }
\vspace{-4mm}
\caption[]{$|F_1(s)|$ in the three--channel case. The thick solid and
long--dashed lines correspond to the unitarised $\chi$PT fits of eqs.~(6.10)
and (6.11) of \cite{JOP:00a} respectively. For comparison, as the thin lines
we have also displayed the corresponding two--channel form factors for (6.10)
and (6.11).
\label{fig9}}
\end{figure}

\begin{figure}[htb]
\centerline{
\rotate[r]{
\epsfysize=14cm\epsffile{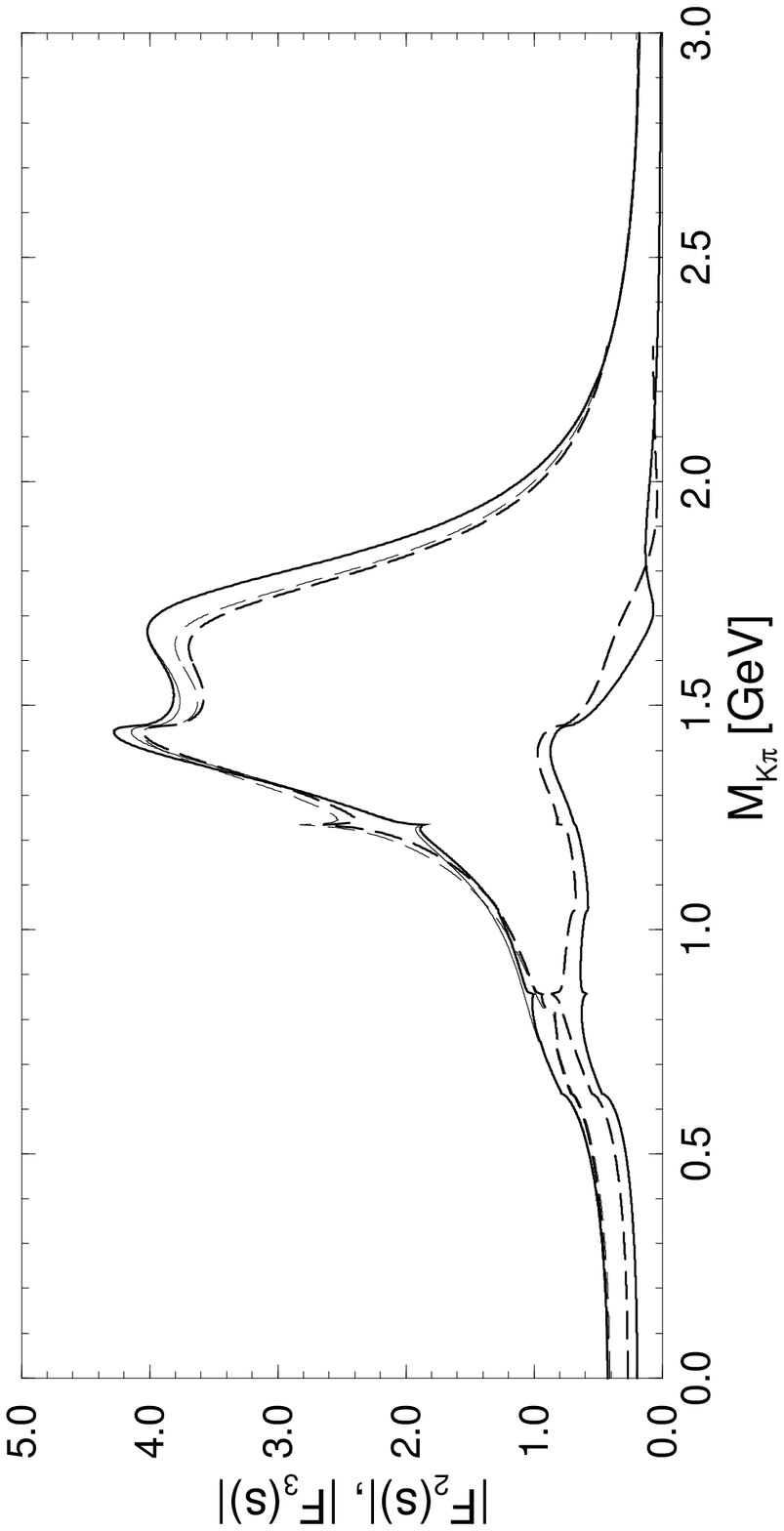} } }
\vspace{-4mm}
\caption[]{$|F_2(s)|$ and $|F_3(s)|$ in the three--channel case. The solid and
long--dashed lines correspond to the unitarised $\chi$PT fits of eqs.~(6.10)
and (6.11) of \cite{JOP:00a} respectively. The two lower lines correspond to
$|F_2(s)|$. For comparison, as the thin lines we have also displayed the
corresponding two--channel form factors for (6.10) and (6.11).
\label{fig10}}
\end{figure}

In figure~\ref{fig9}, we display the absolute values of the form factor
$F_1(s)$ for the three--channel fits (6.10) and (6.11) of ref. \cite{JOP:00a},
and in figure~\ref{fig10} the corresponding form factors $|F_2(s)|$ and
$|F_3(s)|$ are shown. Like in section 4.2, the notation of the lines is (6.10)
solid line and (6.11) long-dashed line. The three--channel form factors are
plotted as thick lines and for comparison, in addition, as the thin lines we
have displayed the two--channel form factors for the fits (6.10) and (6.11).
In figure \ref{fig10}, the two lower lines correspond to $|F_2(s)|$. As can be
observed from this figure, the form factors $F_2(s)$ turns out much smaller
than $F_3(s)$, and for the scalar spectral function, further discussed in
\cite{JOP:01}, it will be completely negligible. Furthermore, the two-- and
three--channel form factors are found to be very similar which again supports
the finding that the $K\eta$ channel is rather unimportant.

Like in the case of the fits (6.10) and (6.11) for two channels, there is
only one solution to the dispersion relations and we just have the freedom to
fix the normalisation $F_1(0)$. The form factors $F_2(0)$ and $F_3(0)$ are
then determined from the dispersion relation. The results for our three--channel
fits are given by:
\begin{eqnarray}
\label{F20F30}
F_2^{(6.10)}(0) &\!=\!& -\,0.195 \,, \quad\qquad
F_3^{(6.10)}(0)  \;=\;     0.426 \,, \\
F_2^{(6.11)}(0) &\!=\!& -\,0.269 \,, \quad\qquad
F_3^{(6.11)}(0)  \;=\;     0.424 \nn \,.
\end{eqnarray}
As can already be guessed from the similarity of the two-- and three--channel
form factors, $F_3(0)$ for the fits (6.10) and (6.11) comes out very close to
the corresponding numbers in the two--channel case. The result for $F_2(0)$
should be compared with the tree--level expectation from ${\rm U}(3)_L\!\times
\!{\rm U}(3)_R$ $\chi$PT in the large-$N_c$ limit including resonances,
$F_2^{\chi\mbox{\tiny PT}}(0) = -\,f_{K\eta}(0)/3 = -\,0.109$. Here for (6.11)
the deviation amounts to more than a factor of two, and for (6.10) it is
somewhat smaller. However, this difference is similar to that of $F_3(0)$
before improving the high energy tail of the experimental data above
$1.75\;\gev$, and hence, until such an improvement is performed for the
three-channel case, one cannot draw a definite conclusion on $F_2(0)$.

Like in the two--channel case, also for the three--channel fits (6.10) and
(6.11) we investigate the low-energy constraints $F_1(\DeKP)$, as well as
the derivative of $F_1(s)$ at zero $D_1(0)$. The results are found as:
\begin{eqnarray}
\label{F1DeKPD10}
F_1^{(6.10)}(\DeKP) &\!=\!& 1.266 \,, \quad\qquad
D_1^{(6.10)}(0)      \;=\;  0.927 \,, \\
F_1^{(6.11)}(\DeKP) &\!=\!& 1.264 \,, \quad\qquad
D_1^{(6.11)}(0)      \;=\;  0.909 \nn \,.
\end{eqnarray}
Again, $F_1(\DeKP)$ for three channels turns out very close to the two--channel
case where also $F_1(\DeKP)$ has been found too large, so that the corresponding
form factors cannot be considered physical. Nevertheless, the finding that for
both fits $F_1(\DeKP)$ is very close is reflected in the fact that the full
form factors $F_1(s)$ turn out very similar. The derivative of $F_1(s)$ at
zero $D_1(0)$ for (6.10) and (6.11), on the other hand, also is close to the
corresponding numbers in the two--channel case, as well as to the $\chi$PT
expectation.

\newsection{Conclusions}

In this work, we have investigated the strangeness--changing scalar form
factors for the $K\pi$, $K\eta$ and $K\eta'$ systems up to $2\;\gev$. To this
end, first theoretical expressions for the scalar form factors were derived in
the framework of the chiral Lagrangian including resonances in the limit of a
large number of colours. In a second step, analyticity and unitarity constraints
were employed to write down a set of dispersion relations for the form factors.
Taking into account previous results on S--wave $K\pi$ scattering
\cite{JOP:00a}, the scalar $K\pi$, $K\eta$ and $K\eta'$ form factors could be
extracted from the solutions of the coupled--channel dispersion relations.

The $\eta$ and $\eta'$ mesons can be described in the ${\rm U}(3)_L\!\times\!
{\rm U}(3)_R$ effective theory in the large-$N_c$ limit. In this framework,
theoretical expressions for the tree--level scalar form factors have been
derived in section~2. However, the couplings of the scalar meson resonances to
the pseudoscalar mesons, $c_d$ and $c_m$, which appear in the resonance chiral
Lagrangian are not very well know. Assuming the scalar form factors to vanish
at infinity, constraints on the scalar couplings could be obtained which are
compatible with present phenomenological knowledge on $c_d$ and $c_m$. In
particular, if the chiral low-energy constants $L_5$ and $L_8$ are saturated
by the scalar meson resonances, we deduced the relation $L_5=2L_8$, independent
of the number of resonances. Further consequences of the theoretical constraints
have already been investigated in our previous analysis of S--wave $K\pi$
scattering \cite{JOP:00a}, and in more detail in section~2.2. Furthermore, in
the chiral framework it was found that the $K\eta$ form factor is strongly
suppressed compared to the $K\pi$ and $K\eta'$ form factors.

Using unitarity and analyticity, we were able to write down a set of coupled
dispersion relations, which relate the scalar form factors to the S--wave,
isospin $1/2$ scattering amplitude. In the elastic, single--channel case,
an analytical solution of the dispersion relation due to Omn\`es is known,
which only depends on the elastic S--wave phase shift. In the coupled--channel
case the set of integral equations had to be solved numerically. To perform
this step, an efficient algorithm to perform the Cauchy principal value
integrals was developed, which is described in appendix~C.2.

Using our results for different fits to the elastic $K\pi$ scattering data
\cite{JOP:00a}, it was found that the scalar $K\pi$ form factor $F_1(s)$ is
not well determined in the single channel case and that higher energy data,
as well as inelastic channels have to be included for a better description.
Since it was shown that the $K\eta$ channel is suppressed, we have concentrated
our detailed investigation of the form factors on the two--channel case with
$K\pi$ and $K\eta'$. In general, the two--channel dispersion relations admit
two linearly independent solutions, and thus we need two integration constants
to fix them unambiguously.

An obvious choice for the first constant is $F_1(0)$ which is very well known
from next-to-leading order $\chi$PT. Different possible choices for the second
constant could be $F_3(0)$, the derivative of $F_1(s)$ at zero $D_1(0)$, or
$F_1(\DeKP)$ at the so-called Callan-Treiman point. We have used the last
constraint because in our opinion it is known most precisely also from
next-to-leading order $\chi$PT. With this information we then deduced
$F_3(0)=0.58\pm0.04$, compatible with the tree--level $\chi$PT expectation in
the large-$N_c$ limit. Furthermore, $F_3(0)$ is also in excellent agreement to
the value from those fits with an improved high energy behaviour that gave rise
to only one independent solution of the form factors.

The three--channel case was used to corroborate that indeed the $K\eta$ form
factor $F_2(s)$ turns out much smaller than $F_1(s)$ and $F_3(s)$, and that the
scalar $K\pi$ and $K\eta'$ form factors were practically unchanged compared to
the two--channel case. This gives us confidence that the more elaborate
two--channel analysis is sufficient for our purpose of determining the
strangeness--changing scalar spectral function since the $K\eta$ contribution
is negligible.

In a forthcoming work \cite{JOP:01}, our results for the scalar $K\pi$ and
$K\eta'$ form factors will be utilised to explicitly calculate the
corresponding strangeness--changing scalar spectral function up to $2\;\gev$.
This spectral function is a key ingredient in the determination of the strange
quark mass from QCD sum rules for the scalar correlator and it is also
important in the corresponding analysis of the Cabibbo--suppressed hadronic
$\tau$ decay width.

\bigskip
\subsection*{Acknowledgements}
This work has been supported in part by the German--Spanish Cooperation
Agreement HA97-0061, by the European Union TMR Network EURODAPHNE
(ERBFMX-CT98-0169), and by DGESIC (Spain) under the grant no. PB97-1261
and DGICYT contract no. BFM2000-1326. M.J. would like to thank the Deutsche
Forschungsgemeinschaft for support.

\newpage
\appendix{\LARGE\bf Appendices}

\newsection{Loop functions}

For completeness, we tabulate here the one-loop function $\bar{J}_{PQ}(s)$
\cite{GL:85} appearing in the next-to-leading order chiral form factors of
eqs.~\eqn{ffxpt} and \eqn{ffeta_xpt}:
\begin{eqnarray}
\bar{J}_{PQ}(s) &\equiv & -{1\over 16\pi^2}\,\int\limits_0^1 \!dx \,
\ln{\left[ {M_P^2 - s x (1-x) - \Delta_{PQ} x \over 
   M_P^2  - \Delta_{PQ} x} \right]} \\
\smvs
& = &
{1\over 32\pi^2}\, \left\{ 2 + 
  \left( {\Delta_{PQ} \over s} - {\Sigma_{PQ}\over \Delta_{PQ}}\right)
  \ln{\left({M_Q^2\over M_P^2}\right)} - {\lambda_{PQ}\over s}
  \ln{\left[ {\left( s+\lambda_{PQ}\right)^2 - \Delta_{PQ}^2\over
     \left( s-\lambda_{PQ}\right)^2 - \Delta_{PQ}^2}\right]} \right\} \,, \nn
\end{eqnarray}
with
$$
\Sigma_{PQ} \equiv M_P^2 + M_Q^2  \; ; \quad
\Delta_{PQ} \equiv M_P^2 - M_Q^2  \; ; \quad
\lambda_{PQ}^2\equiv  \left[ s - \left( M_P+M_Q\right)^2\right]
  \left[ s - \left( M_P-M_Q\right)^2\right] \; .
$$

\newsection{Three--channel dispersion relations}

Below we present the dispersion relations for the scalar form factors
$F_k(s)$ in the full three--channel case including $K\pi$, $K\eta$ and
$K\eta'$:

\begin{eqnarray}
\label{3in}
F_1(s) &=& \frac{1}{\pi}\int\limits_{s_{th\,1}}^\infty \!\!ds'\,
\frac{\sigma_1(s') F_1(s')\,t_0^{11}(s')^*}{(s'-s-i0)} +
\frac{1}{\pi}\int\limits_{s_{th\,2}}^\infty \!\!ds'\,
\frac{\sigma_2(s') F_2(s')\,t_0^{12}(s')^*}{(s'-s-i0)} \nn \\
\smvs
&+& \frac{1}{\pi}\int\limits_{s_{th\,3}}^\infty \!\!ds'\,
\frac{\sigma_3(s') F_3(s')\,t_0^{13}(s')^*}{(s'-s-i0)} \;, \nn \\
\bmvs
F_2(s) &=& \frac{1}{\pi}\int\limits_{s_{th\,1}}^\infty \!\!ds'\,
\frac{\sigma_1(s') F_1(s')\,t_0^{12}(s')^*}{(s'-s-i0)} +
\frac{1}{\pi}\int\limits_{s_{th\,2}}^\infty \!\!ds'\,
\frac{\sigma_2(s') F_2(s')\,t_0^{22}(s')^*}{(s'-s-i0)} \nn \\
\smvs
&+& \frac{1}{\pi}\int\limits_{s_{th\,3}}^\infty \!\!ds'\,
\frac{\sigma_3(s') F_3(s')\,t_0^{23}(s')^*}{(s'-s-i0)} \;, \\
\bmvs
F_3(s) &=& \frac{1}{\pi}\int\limits_{s_{th\,1}}^\infty \!\!ds'\,
\frac{\sigma_1(s') F_1(s')\,t_0^{13}(s')^*}{(s'-s-i0)} +
\frac{1}{\pi}\int\limits_{s_{th\,2}}^\infty \!\!ds'\,
\frac{\sigma_2(s') F_2(s')\,t_0^{23}(s')^*}{(s'-s-i0)} \nn \\
\smvs
&+& \frac{1}{\pi}\int\limits_{s_{th\,3}}^\infty \!\!ds'\,
\frac{\sigma_3(s') F_3(s')\,t_0^{33}(s')^*}{(s'-s-i0)} \;. \nn
\end{eqnarray}

\newsection{Numerical solution of the dispersion relation}

To illustrate our method for the numerical solution of the dispersion relation,
below we shall discuss the single channel case. The coupled channel case can be
treated in a completely analogous fashion.

\subsection{Single channel case}

In the single channel case, the dispersion relation reads:
\begin{equation}
\label{di}
\RE F_1(s) \; = \;
\cP\!\!\int\limits_{s_{th}}^\infty \!\!ds'\,\frac{f_1(s')}{(s'-s)} \; = \;
\cP\!\!\!\int\limits_{s_{th}}^{s_{cut}} \!\!ds'\,\frac{f_1(s')}{(s'-s)} +
\cP\!\!\!\int\limits_{s_{cut}}^\infty \!\!ds'\,\frac{f_1(s')}{(s'-s)} \,,
\end{equation}
with
\begin{equation}
\label{uc}
f_1(s') \; \equiv \; {1\over\pi}\,\IM F_1(s') \; = \;
{1\over\pi}\,\sigma_1(s')F_1(s')\,t_0^{11}(s')^* \,,
\end{equation}
and $s_{th}=(M_K+M_\pi)^2$. For numerical purposes on the right-hand side
we have split the integration range at an energy $s_{cut}$. Equation \eqn{di}
can also be written as a once subtracted dispersion relation:
\begin{equation}
\label{disu}
\RE F_1(s) \; = \; \int\limits_{s_{th}}^\infty \!\!ds'\,\frac{f_1(s')}{s'}
+ \cP\!\!\int\limits_{s_{th}}^\infty \!\!ds'\,\frac{sf_1(s')}
{s'(s'-s)} \,.
\end{equation}
Assuming that $f_1(s)$ vanishes for $s\to\infty$, such that the first integral
converges, it can be identified with $F_1(0)$. In the numerical analysis, we
have also investigated solving the subtracted dispersion relation imposing
$F_1(0)$ as a boundary condition.

For the numerical integration, it is convenient to perform a change of
variables such that the integration range becomes finite. We shall transform
both energy intervals to the interval $(0,1)$. For the low-energy interval
$(s_{th},s_{cut})$ a convenient transformation is:
\begin{equation}
s' \; \equiv \; s_{th} + (s_{cut}-s_{th})\,x'^2 \,; \qquad
x' \; = \; \sqrt{\frac{(s'-s_{th})}{(s_{cut}-s_{th})}} \,,
\end{equation}
whereas for the high-energy region $(s_{cut},\infty)$ we chose:
\begin{equation}
s' \; \equiv \; s_{th}\,\frac{(1-b\,z')}{(1-z')} \;; \qquad
z' \; = \; \frac{(s'-s_{th})}{(s'-b\,s_{th})} \,.
\end{equation}
The additional parameter $b<1$ in the second transformation has been introduced
to allow for increasing the convergence of the method or to check the stability
of the solution. Defining the additional variables $x\equiv\sqrt{(s-s_{th})/
(s_{cut}-s_{th})}$ and $z\equiv (s-s_{th})/(s-b\,s_{th})$, the unsubtracted
dispersion relation takes the form:
\begin{equation}
\label{dix}
\RE F_1(x) \; = \;
\cP\!\!\int\limits_0^1 \! dx'\,\frac{2x'f_1(x')}{(x'^2-x^2)} +
\cP\!\!\int\limits_0^1 \! dz'\,\frac{(1-z)f_1(z')}{(1-z')(z'-z)} \,,
\end{equation}
whereas the subtracted dispersion relation is given by:
\begin{equation}
\label{disx}
\RE F_1(x) \; = \; F_1(0) +
\cP\!\!\int\limits_0^1 \! dx'\,\frac{2x'[1+(a-1)x^2]f_1(x')}
{[1+(a-1)x'^2](x'^2-x^2)} +
\cP\!\!\int\limits_0^1 \! dz'\,\frac{(1-bz)f_1(z')}{(1-bz')(z'-z)} \,,
\end{equation}
with $a\equiv s_{cut}/s_{th}$. The additional factor of $x'$ in the low-energy
integral has the advantage of smoothing out the square-root singularity
$x=x'=0$.

As a further ingredient, for the iteration of the dispersion relation, we
require a reasonably accurate numerical integration routine for the Cauchy
principal value integrals. Generally, the most accurate integration routines
are based on the Gauss algorithm. In the next section, we shall thus develop
a Gauss algorithm for the Cauchy kernel.

\subsection{Gauss quadrature for Cauchy kernel}

Gaussian quadrature algorithms can be designed such that the approximation
\begin{equation}
\int\limits_a^b f(y)w(y)dy \; \approx \; \sum\limits_{j=1}^N \,w_j f(y_j)
\end{equation}
is exact if $f(x)$ is a polynomial. The task for finding such algorithms is
finding the set of weights $w_j$ and abscissas $y_j$ to accomplish this feat,
given a weight function $w(y)$.

The starting point for developing the Gauss algorithm lies in finding a
complete basis of functions which are orthogonal with respect to the weight
function $w(y)$ \cite{numrec}. For the trivial weight $w(y)=1$, on the interval
$(-1,1)$, this set of functions can be chosen to be the Legendre polynomials
$P_n(y)$ and the corresponding integration formula is the so-called
Gauss--Legendre algorithm.

Defining the scalar product of two functions $f$ and $g$ over the weight
function $w(y)$ as
\begin{equation}
\langle f|g\rangle \; = \; \int\limits_a^b f(y)g(y)w(y)dy \,,
\end{equation}
we therefore seek functions $u_n(y)$ such that $\langle u_m|u_n\rangle$
is zero if $m\neq n$. Choosing the Cauchy weight $w(y)=1/(x-y)$, it can be
shown that again on the interval $(-1,1)$ a set of orthogonal functions is
given by
\begin{equation}
u_n(x,y) \; = \; P_n(y) - \frac{Q_n(x)}{Q_{n-1}(x)}\,P_{n-1}(y) \,,
\end{equation}
where $Q_n(x)$ are the associated Legendre functions of the second kind
\cite{gr:80}.  The normalisation is easily found to be:
\begin{equation}
\langle u_n|u_n\rangle \; = \; 2\,Q_n(x)\biggl[\, P_n(x) -
\frac{Q_n(x)}{Q_{n-1}(x)}\,P_{n-1}(x) \,\biggr] \; = \;
2\,Q_n(x)\,u(x,x) \,.
\end{equation}

For a Gauss algorithm of order $N$, the set of abscissas is now given by the
zeros $y_j$ of the function $u_N(y)$ in the interval $(a,b)$, and the weights
$w_j$ can be calculated from a general formula, which can for example be found
in the Numerical Recipes \cite{numrec}. In our particular case, however, we
still have the complication that the weight function, and thus also the $u_n$,
also depend on the additional variable $x$. The problem simplifies considerably,
if we only evaluate our integral at the zeros $x_i$ of the function $Q_N(x)$.
Then our set of orthogonal functions is again the Legendre polynomials, and
from a straightforward calculation, the weights $w_j(x_i)$ are obtained to be
\begin{equation}
w_j(x_i) \; = \; \frac{2(1-y_j^2)}{N(x_i-y_j)}\,\frac{Q_{N-1}(x_i)\,
P_N(x_i)}{P_{N-1}^2(y_j)} \,.
\end{equation}
Our final integration formula reads:
\begin{equation}
\int\limits_{-1}^1 f(y)\,\frac{dy}{(x_i-y)} \; = \;
\sum\limits_{j=1}^N \,w_j(x_i) f(y_j) \,.
\end{equation}

It is a trivial matter to find the integration formula for a general interval
$(a,b)$ using linear transformations. The result for the integral at values
of $x$ different from $x_i$ can be calculated with the help of standard
interpolation algorithms \cite{numrec}. A final comment concerns the practical
calculation of the roots of $Q_N$. Once the $N$ roots of $P_N$ on the interval
$(-1,1)$ are computed with a standard algorithm \cite{numrec}, the $N+1$
roots of $Q_N$ are easily searched for, because the always interleave the
roots of $P_N$.

Although the presented Gauss algorithm is rather simple, we were unable to
find it in the literature. Thus, in appendix~D we present our Fortran
implementation of the algorithm which is based on the Gauss--Legendre algorithm
given in the Numerical Recipes \cite{numrec}.

\newpage

\newsection{Fortran code for the Gauss routine}

\begin{mathematica}

C----------------------------------------------------------------------
      !
C     ! Small program to test the subroutine "cauleg"
C     ! Calculates the integral int( sin(y)^2/(x-y), y=0:3);
C     ! Last change: 12.1.2001
C     ! Matthias Jamin: m.jamin@thphys.uni-heidelberg.de
      !
      Program CauLegTest
      Implicit Double Precision (A-z)
      Integer i, j, ngau
      Parameter (ngau = 63, a = 0.D0, b = 3.D0)
      Dimension x(ngau+1), y(ngau), w(ngau+1,ngau)
      !
C     ! Calculate zeros of the Legendre polynomials and Cauchy weights
      !
      Call cauleg(a,b,x,y,w,ngau)
      !
C     ! Calculate integral at the zeros x(i) of Q_n(x)
      !
      Open (unit=8, file='cauleg.out')
      Do 20 i=1,ngau+1
      !
      sum = 0.D0
      Do 10 j=1,ngau
   10 sum = sum + dsin(y(j))**2*w(i,j)
      si1 = sum
      !
C     ! With the CERN library "mathlib" this is the exact result
      !
      xx = x(i)
      si2 = ( dcos(2.D0*xx)*(dcosin(2.D0*(a-xx))-dcosin(2.D0*(b-xx)))
     .      + dsin(2.D0*xx)*(dsinin(2.D0*(xx-a))-dsinin(2.D0*(xx-b)))
     .      + dlog((b-xx)/(xx-a)) )/2.D0
      !
   20 Write (8,*) x(i), si1, si2
      Close (8)
      !
      Return
      End
C-----------------------------------------------------------------------
      !
C     ! Gauss routine for integration with the Cauchy kernel 1/(x-y)
C     ! based on the routine "gauleg" from the Numerical Recipes.
C     ! Uses legp(x,n) and legq(x,n)
      !
C     ! Given the lower and upper limits of integration x1 and x2, and
C     ! given n, the routine returns arrays x(1:n+1), y(1:n), w(1:n+1,1:n)
C     ! containing the points x where the integral should be evaluated,
C     ! as well as the abscissas and the weights of the Gauss-Legendre
C     ! n-point quadrature formula. x(1:n+1) and y(1:n) contain the roots
C     ! of the Legendre functions Q_n and P_n of order n, shifted to the
C     ! interval (x1,x2). Other values of x have to be calculated with
C     ! standard interpolation methods.
      !
      Subroutine cauleg(x1,x2,x,y,w,n)
      Implicit Double Precision (A-z)
      Integer i, j, m, n
      Dimension x(n+1), y(n), w(n+1,n)
      Parameter (Pi = 3.141592653589793238D0, eps = 1.D-15)
      !
      m = (n+1)/2
      xm = 0.5D0*(x2+x1)
      xl = 0.5D0*(x2-x1)
      !
      Do 20 i=1,m
         z = dcos(Pi*(i-0.25D0)/(n+0.5D0))
   10    z1 = z
         Call legp(p1,pp,z,n)
         z = z1-p1/pp
         If (dabs(z-z1).gt.eps) Goto 10
         y(i) =     xm-xl*z
         y(n-i+1) = xm+xl*z
   20 Continue
      !
      Do 40 i=1,m+1
         If (i.eq.1) Then
            z = -1.D0+1.D-9
         Else
            z = 0.5D0*(y(i)+y(i-1)-2.D0*xm)/xl
         End If
   30    z1 = z
         Call legq(q1,qp,z,n)
         z = z1-q1/qp
         If (dabs(z-z1).gt.eps) Goto 30
         x(i) =     xm+xl*z
         x(n-i+2) = xm-xl*z
   40 Continue
      !
      Do 50 i=1,m+1
         xi = (x(i)-xm)/xl
         Call legp(p1,pp,xi,n)
         Call legq(q2,qp,xi,n-1)
         Do 50 j=1,n
            yj = (y(j)-xm)/xl
            Call legp(p2,pp,yj,n-1)
            w(i,j) = 2.D0/dble(n)*(1.D0-yj*yj)
     .               /(yj-xi)*p1*q2/p2/p2
            w(n-i+2,n-j+1) = -w(i,j)
   50 Continue
      !
      Return
      End
C-----------------------------------------------------------------------
      !
C     ! Calculates the Legendre polynomials P_n(x) and P'_n(x)
C     ! P_n(x) is in p1 and P'_n(x) in pp.
      !
      Subroutine legp(p1,pp,x,n)
      Implicit Double Precision (A-z)
      Integer j, n
      !
      p2 = 1.D0
      p1 = x
      Do 10 j=2,n
         p3 = p2
         p2 = p1
         p1 = ((2.D0*j-1.D0)*x*p2-(j-1.D0)*p3)/j
   10 Continue
      pp = n*(x*p1-p2)/(x*x-1.D0)
      !
      Return
      End
C-----------------------------------------------------------------------
      !
C     ! Calculates the Legendre functions Q_n(x) and Q'_n(x).
C     ! Q_n(x) is in q1 and Q'_n(x) in qp.
      !
      Subroutine legq(q1,qp,x,n)
      Implicit Double Precision (A-z)
      Integer j, n
      !
      q2 = 0.5D0*dlog(dabs((1.D0+x)/(1.D0-x)))
      q1 = x*q2-1.D0
      Do 10 j=2,n
         q3 = q2
         q2 = q1
         q1 = ((2.D0*j-1.D0)*x*q2-(j-1.D0)*q3)/j
   10 Continue
      qp = n*(x*q1-q2)/(x*x-1.D0)
      !
      Return
      End
C-----------------------------------------------------------------------

\end{mathematica}

\newpage

\begin{thebibliography}{10}

\bibitem{JOP:00a}
{\sc M.~Jamin}, {\sc J.~A. Oller}, and {\sc A.~Pich},
\newblock S-wave $K\pi$ scattering in chiral perturbation theory with
  resonances,
\newblock {\em Nucl. Phys.} {\bf B587} (2000) 331.

\bibitem{WE:79}
{\sc S.~Weinberg},
\newblock Phenomenological Lagrangians,
\newblock {\em Physica} {\bf A96} (1979) 327.

\bibitem{GL:84}
{\sc J.~Gasser} and {\sc H.~Leutwyler},
\newblock Chiral perturbation theory to one loop,
\newblock {\em Ann. Phys.} {\bf 158} (1984) 142.

\bibitem{GL:85}
{\sc J.~Gasser} and {\sc H.~Leutwyler},
\newblock Chiral perturbation theory: expansions in the mass of the strange
  quark,
\newblock {\em Nucl. Phys.} {\bf B250} (1985) 465, 517, 539.

\bibitem{EC:95}
{\sc G.~Ecker},
\newblock Low-energy QCD,
\newblock {\em Prog. Part. Nucl. Phys.} {\bf 36} (1996) 71.

\bibitem{PI:95}
{\sc A.~Pich},
\newblock Chiral perturbation theory,
\newblock {\em Rept. Prog. Phys.} {\bf 58} (1995) 563.

\bibitem{ME:93}
{\sc U.-G. Mei{\ss}ner},
\newblock Recent developments in chiral perturbation theory,
\newblock {\em Rept. Prog. Phys.} {\bf 56} (1993) 903.

\bibitem{EGPR:89}
{\sc G.~Ecker}, {\sc J.~Gasser}, {\sc A.~Pich}, and {\sc E.~{de Rafael}},
\newblock The role of resonances in chiral perturbation theory,
\newblock {\em Nucl. Phys.} {\bf B321} (1989) 311.

\bibitem{EGLPR:89}
{\sc G.~Ecker}, {\sc J.~Gasser}, {\sc H.~Leutwyler}, {\sc A.~Pich}, and {\sc
  E.~{de Rafael}},
\newblock Chiral Lagrangians for massive spin 1 fields,
\newblock {\em Phys. Lett.} {\bf B223} (1989) 425.

\bibitem{TH:74}
{\sc G.~'t~Hooft},
\newblock A planar diagram theory for strong interactions,
\newblock {\em Nucl. Phys.} {\bf B72} (1974) 461.

\bibitem{WI:79}
{\sc E.~Witten},
\newblock Baryons in the $1/N_c$ expansion,
\newblock {\em Nucl. Phys.} {\bf B160} (1979) 57.

\bibitem{PPR:98}
{\sc S.~Peris}, {\sc M.~Perrottet}, and {\sc E.~{de Rafael}},
\newblock Matching long and short distances in large-$N_c$ QCD,
\newblock {\em JHEP} {\bf 05} (1998) 011.

\bibitem{NPRT:83}
{\sc S.~Narison}, {\sc N.~Paver}, {\sc E.~{de Rafael}}, and {\sc D.~Treleani},
\newblock Light quark mass differences in quantum chromodynamics,
\newblock {\em Nucl. Phys.} {\bf B212} (1983) 365.

\bibitem{JM:95}
{\sc M.~Jamin} and {\sc M.~M{\"u}nz},
\newblock The Strange quark mass from QCD sum rules,
\newblock {\em Z. Phys.} {\bf C66} (1995) 633.

\bibitem{ChDPS:95}
{\sc K.~G. Chetyrkin}, {\sc C.~A. Dominguez}, {\sc D.~Pirjol}, and {\sc
  K.~Schilcher},
\newblock Mass singularities in light quark correlators: the strange quark
  case,
\newblock {\em Phys. Rev.} {\bf D51} (1995) 5090.

\bibitem{cps:96}
{\sc K.~G. Chetyrkin}, {\sc D.~Pirjol}, and {\sc K.~Schilcher},
\newblock Order $\alpha_s^3$ determination of the strange quark mass,
\newblock {\em Phys. Lett.} {\bf B404} (1997) 337.

\bibitem{CFNP:97}
{\sc P.~Colangelo}, {\sc F.~{de Fazio}}, {\sc G.~Nardulli}, and {\sc N.~Paver},
\newblock On the QCD sum rule determination of the strange quark mass,
\newblock {\em Phys. Lett.} {\bf B408} (1997) 340.

\bibitem{JA:98}
{\sc M.~Jamin},
\newblock The strange quark mass from scalar sum rules updated,
\newblock {\em Nucl. Phys. Proc. Suppl.} {\bf 64} (1998) 250,
\newblock Proc. of {\em QCD 97}, Montpellier, July 1997.

\bibitem{BGM:98}
{\sc T.~Bhattacharya}, {\sc R.~Gupta}, and {\sc K.~Maltman},
\newblock Duality and the extraction of light quark masses from finite energy
  and QCD sum rules,
\newblock {\em Phys. Rev.} {\bf D57} (1998) 5455.

\bibitem{mal:99}
{\sc K.~Maltman},
\newblock The strange quark mass from finite energy sum rules,
\newblock {\em Phys. Lett.} {\bf B462} (1999) 195.

\bibitem{PP:98}
{\sc A.~Pich} and {\sc J.~Prades},
\newblock Perturbative quark mass corrections to the $\tau$ hadronic width,
\newblock {\em JHEP} {\bf 06} (1998) 013.

\bibitem{ALEPH:99}
{\sc {ALEPH collaboration}},
\newblock Study of $\tau$ decays involving kaons, spectral functions and
  determination of the strange quark mass,
\newblock {\em Eur. Phys. J.} {\bf C11} (1999) 599.

\bibitem{PP:99}
{\sc A.~Pich} and {\sc J.~Prades},
\newblock Strange quark mass determination from Cabibbo-suppressed $\tau$
  decays,
\newblock {\em JHEP} {\bf 10} (1999) 004.

\bibitem{kkp:00}
{\sc J.~G. K{\"o}rner}, {\sc F.~Krajewski}, and {\sc A.~A. Pivovarov},
\newblock Determination of the strange quark mass from Cabibbo suppressed tau
  decays with resummed perturbation theory in an effective scheme,
\newblock {\em Eur. Phys. J.} {\bf C20} (2001) 259.

\bibitem{KM:00}
{\sc J.~Kambor} and {\sc K.~Maltman},
\newblock The strange quark mass from flavor breaking in hadronic $\tau$
  decays,
\newblock {\em Phys. Rev.} {\bf D62} (2000) 093023.

\bibitem{ChDGHPP:01}
{\sc S.~Chen}, {\sc M.~Davier}, {\sc E.~G\'amiz}, {\sc A.~H{\"o}cker}, {\sc
  A.~Pich}, {\em et~al.},
\newblock Strange quark mass from the invariant mass distribution of
  Cabibbo-suppressed $\tau$ decays,
\newblock (2001),
\newblock hep-ph/0105253, to appear in {\em Eur. Phys. J}.

\bibitem{tru:88}
{\sc T.~N. Truong},
\newblock Chiral perturbation theory and final state theorem,
\newblock {\em Phys. Rev. Lett.} {\bf 61} (1988) 2526.

\bibitem{dht:90}
{\sc A.~Dobado}, {\sc M.~J. Herrero}, and {\sc T.~N. Truong},
\newblock Unitarized chiral perturbation theory and elastic pion-pion
  scattering,
\newblock {\em Phys. Lett.} {\bf B235} (1990) 134.

\bibitem{oo:97}
{\sc J.~A. Oller} and {\sc E.~Oset},
\newblock Chiral symmetry amplitudes in the S-wave isoscalar and isovector
  channels and the $\sigma$, $f_0(980)$, $a_0(980)$ scalar mesons,
\newblock {\em Nucl. Phys. \rm {\bf A620} (1997) 438, Erratum \em Nucl. Phys.}
  {\bf A652} (1999) 407.

\bibitem{GP:97}
{\sc F.~Guerrero} and {\sc A.~Pich},
\newblock Effective field theory description of the pion form factor,
\newblock {\em Phys. Lett.} {\bf B412} (1997) 382.

\bibitem{GU:98}
{\sc F.~Guerrero},
\newblock Study of the resummation of chiral logarithms in the exponentiated
  expression for the pion form factor,
\newblock {\em Phys. Rev.} {\bf D57} (1998) 4136.

\bibitem{OOP:99}
{\sc J.~A. Oller}, {\sc E.~Oset}, and {\sc J.~R. Pel{\'a}ez},
\newblock Meson-meson and meson-baryon interactions in a chiral
  non-perturbative approach,
\newblock {\em Phys. Rev. Lett. \rm {\bf 80} (1998) 3452, \em Phys. Rev.} {\bf
  D59} (1999) 074001.

\bibitem{OO:99}
{\sc J.~A. Oller} and {\sc E.~Oset},
\newblock N/D description of two-meson amplitudes and chiral symmetry,
\newblock {\em Phys. Rev.} {\bf D60} (1999) 074023.

\bibitem{OOR:99}
{\sc J.~A. Oller}, {\sc E.~Oset}, and {\sc A.~Ramos},
\newblock Chiral unitary approach to meson-meson and meson-baryon interactions
  and nuclear applications,
\newblock {\em Prog. Part. Nucl. Phys.} {\bf 45} (2000) 157.

\bibitem{dpp:00}
{\sc D.~{G\'omez Dumm}}, {\sc A.~Pich}, and {\sc J.~Portoles},
\newblock The hadronic off-shell width of meson resonances,
\newblock {\em Phys. Rev.} {\bf D62} (2000) 054014.

\bibitem{mo:00}
{\sc U.-G. Mei{\ss}ner} and {\sc J.~A. Oller},
\newblock $J/\psi\to\phi\pi\pi$ ($K\overline K$) decays, chiral dynamics and
  OZI violation,
\newblock {\em Nucl. Phys.} {\bf A679} (2001) 671.

\bibitem{oop:00}
{\sc J.~A. Oller}, {\sc E.~Oset}, and {\sc J.~E. Palomar},
\newblock Pion and kaon vector form factors,
\newblock {\em Phys. Rev.} {\bf D63} (2001) 114009.

\bibitem{pp:01}
{\sc A.~Pich} and {\sc J.~Portoles},
\newblock The vector form factor of the pion from unitarity and analyticity: A
  model-independent approach,
\newblock {\em Phys. Rev.} {\bf D63} (2001) 093005.

\bibitem{oll:00}
{\sc J.~A. Oller},
\newblock The case of a $W W$ dynamical scalar resonance within a chiral
  effective description of the strongly interacting Higgs sector,
\newblock {\em Phys. Lett.} {\bf B477} (2000) 187.

\bibitem{leu:96}
{\sc H.~Leutwyler},
\newblock Bounds on the light quark masses,
\newblock {\em Phys. Lett.} {\bf B374} (1996) 163.

\bibitem{her:97}
{\sc P.~Herrera-Sikl\'ody}, {\sc J.~I. Latorre}, {\sc P.~Pascual}, and {\sc
  J.~Taron},
\newblock Chiral effective Lagrangian in the large-$N_c$ limit: the nonet case,
\newblock {\em Nucl. Phys.} {\bf B497} (1997) 345.

\bibitem{her:98}
{\sc P.~Herrera-Sikl\'ody}, {\sc J.~I. Latorre}, {\sc P.~Pascual}, and {\sc
  J.~Taron},
\newblock $\eta$--$\eta'$ mixing from ${\rm U}(3)_L\!\times\! {\rm U}(3)_R$
  chiral perturbation theory,
\newblock {\em Phys. Lett.} {\bf B419} (1998) 326.

\bibitem{abt:01}
{\sc G.~Amoros}, {\sc J.~Bijnens}, and {\sc P.~Talavera},
\newblock QCD isospin breaking in meson masses, decay constants and quark mass
  ratios,
\newblock {\em Nucl. Phys.} {\bf B602} (2001) 87.

\bibitem{est:78}
{\sc {P.~Estabrooks~et~al.}},
\newblock Study of $K\pi$ scattering using the reactions $K^\pm p \to
  K^\pm\pi^+ n$ and $K^\pm p \to K^\pm\pi^- \Delta^{++}$ at 13 GeV/c,
\newblock {\em Nucl. Phys.} {\bf B133} (1978) 490.

\bibitem{ast:88}
{\sc {D.~Aston~et~al.}},
\newblock A study of $K^-\pi^+$ scattering in the reaction $K^- p \to K^- \pi^+
  n$ at 11 GeV/c,
\newblock {\em Nucl. Phys.} {\bf B296} (1988) 493.

\bibitem{WA:55}
{\sc K.~M. Watson},
\newblock Some general relations between the photoproduction and scattering of
  $\pi$ mesons,
\newblock {\em Phys. Rev.} {\bf 95} (1954) 228.

\bibitem{babelon}
{\sc O.~Babelon}, {\sc J.-L. Basdevant}, {\sc D.~Caillerie}, and {\sc
  G.~Mennessier},
\newblock Unitarity and inelastic final state interactions,
\newblock {\em Nucl. Phys.} {\bf B113} (1976) 445.

\bibitem{om:58}
{\sc R.~Omn\`es},
\newblock On the Solution of certain singular integral equations of quantum
  field theory,
\newblock {\em Nuovo Cim.} {\bf 8} (1958) 316.

\bibitem{PP:00}
{\sc E.~Pallante} and {\sc A.~Pich},
\newblock Final state interactions in kaon decays,
\newblock {\em Nucl. Phys.} {\bf B592} (2001) 294.

\bibitem{dona:90}
{\sc J.~F. Donoghue}, {\sc J.~Gasser}, and {\sc H.~Leutwyler},
\newblock The decay of a light Higgs boson,
\newblock {\em Nucl. Phys.} {\bf B343} (1990) 341.

\bibitem{MS:70}
{\sc A.~D. Martin} and {\sc T.~D. Spearman},
\newblock {\em Elementary Particle Theory},
\newblock North-Holland, Amsterdam, 1970.

\bibitem{JOP:01}
{\sc M.~Jamin}, {\sc J.~A. Oller}, and {\sc A.~Pich},
\newblock Light quark masses from scalar sum rules,
\newblock IFIC/01-52, hep-ph/0110194.

\bibitem{min:94}
{\sc F.~James} and {\sc M.~Roos},
\newblock 'MINUIT' a system for function minimization and analysis of the
  parameter errors and correlations,
\newblock {\em Comput. Phys. Commun.} {\bf 10} (1975) 343.

\bibitem{pdg:00}
{\sc {D.~E.~Groom et al.}},
\newblock Review of particle physics,
\newblock {\em Eur. Phys. J.} {\bf C15} (2000) 1.

\bibitem{numrec}
{\sc {W.~H.~Press~et~al.}},
\newblock {\em Numerical Recipes},
\newblock Cambridge University Press, 1992.

\bibitem{gr:80}
{\sc I.~S. Gradshteyn} and {\sc I.~M. Ryzhik},
\newblock {\em Tables of integrals, series, and products},
\newblock Academic Press, 1980.

\end{thebibliography}

\end{document}